\documentclass{aastex63}

\usepackage{amssymb,amsmath}
\usepackage{braket}
\usepackage{natbib}
\usepackage{hyperref}
\usepackage{array}
\usepackage{afterpage}
\usepackage{graphicx}
\usepackage{bm}
\usepackage{multirow}
\usepackage{lmodern}

\newcommand{\PSUAA}{Department of Astronomy \& Astrophysics, 525 Davey Laboratory, The Pennsylvania State University, University Park, PA, 16802, USA}
\newcommand{\PSUCEHW}{Center for Exoplanets and Habitable Worlds, 525 Davey Laboratory, The Pennsylvania State University, University Park, PA, 16802, USA}
\newcommand{\PSUAS}{Center for Astrostatistics, 525 Davey Lab, The Pennsylvania State University, University Park, PA, 16802, USA}
\newcommand{\PSUICDS}{Institute for Computational and Data Sciences, 525 Davey Lab, The Pennsylvania State University, University Park, PA, 16802, USA}

\newcommand{\Teff}{T_{\mathrm{eff}}}
\newcommand{\logg}{\log{g}}

\newcommand{\logRHK}{\log{R'_{\mathrm{HK}}}}

\newcommand{\ms}{m$\,$s$^{-1}$}
\newcommand{\cms}{cm$\,$s$^{-1}$}

\newcounter{x}
\def\ToRoman#1{\setcounter{x}{#1}\Roman{x}}

\newcommand{\I}{Architecture \ToRoman{1}}
\newcommand{\IIa}{Architecture \ToRoman{2}a}
\newcommand{\IIb}{Architecture \ToRoman{2}b}
\newcommand{\VIIIa}{Architecture \ToRoman{8}a}
\newcommand{\VIIIb}{Architecture \ToRoman{8}b}

\defcitealias{Pereira2019}{P19}
\defcitealias{Guo2022}{Guo22}
\defcitealias{Gilbertson2020}{Gil20b}

\definecolor{jkl}{rgb}{0.4, 0.0, 0.6}

\definecolor{ebf}{rgb}{0.95, 0.5, 0.15}

\begin{document}

\title{Impact of Correlated Noise on the Mass Precision of Earth-analog Planets in Radial Velocity Surveys}

\author[0000-0002-4927-9925]{Jacob K. Luhn}
\affiliation{\PSUAA}
\affiliation{\PSUCEHW}
\affiliation{Department of Physics and Astronomy, The University of California, Irvine, Irvine, CA 92697, USA}

\author{Eric B. Ford}
\affiliation{\PSUAA}
\affiliation{\PSUCEHW}
\affiliation{\PSUAS}
\affiliation{\PSUICDS}
\affiliation{Institute for Advanced Study}

\author{Zhao Guo}
\affiliation{\PSUAA}
\affiliation{\PSUCEHW}
\affiliation{Department of Applied Mathematics and Theoretical Physics, University of Cambridge, Cambridge CB3 0WA, United Kingdom}

\author[0000-0002-1743-3684]{Christian Gilbertson}
\affiliation{\PSUAA}
\affiliation{\PSUCEHW}
\affiliation{\PSUICDS}

\author{Patrick Newman}
\affiliation{Department of Physics and Astronomy, George Mason University, 4400 University Drive, MSN 3F3, Fairfax, VA 22030}

\author[0000-0002-8864-1667]{Peter Plavchan}
\affiliation{Department of Physics and Astronomy, George Mason University, 4400 University Drive, MSN 3F3, Fairfax, VA 22030}

\author[0000-0002-0040-6815]{Jennifer A. Burt}
\affiliation{Jet Propulsion Laboratory, California Institute of Technology, 4800 Oak Grove Drive, Pasadena, CA 91109, USA}

\author{Johanna Teske}
\affiliation{Carnegie Earth and Planets Laboratory, 5241 Broad Branch Road, NW, Washington, DC 20015, USA}

\author[0000-0002-5463-9980]{Arvind F.\ Gupta}
\affiliation{\PSUAA}
\affiliation{\PSUCEHW}

\shorttitle{Correlated Noise in RV Surveys}


\email{jluhn@uci.edu}

\begin{abstract}
Characterizing the masses and orbits of near-Earth-mass planets is crucial for interpreting observations from future direct imaging missions (e.g., HabEx, LUVOIR). 
Therefore, the Exoplanet Science Strategy report \citep{ESS2018} recommended further research so future extremely precise radial velocity surveys could contribute to the discovery and/or characterization of near-Earth-mass planets in the habitable zones of nearby stars prior to the launch of these future imaging missions. 
\citet{EPRVWGreport} simulated such 10-year surveys under various telescope architectures, demonstrating they can precisely measure the masses of potentially habitable Earth-mass planets in the absence of stellar variability. 
Here, we investigate the effect of stellar variability on the signal-to-noise ratio (SNR) of the planet mass measurements in these simulations.  
We find that correlated noise due to active regions has the largest effect on the observed mass SNR, reducing the SNR by a factor of $\sim$5.5 relative to the no-variability scenario --- granulation reduces by a factor of $\sim$3, while p-mode oscillations has little impact on the proposed survey strategies. 
We show that in the presence of correlated noise, 5-\cms{} instrumental precision offers little improvement over 10-\cms{} precision, highlighting the need to mitigate astrophysical variability. 
With our noise models, extending the survey to 15 years doubles the number of Earth-analogs with mass SNR~$ > 10$, and reaching this threshold for any Earth-analog orbiting a star $> 0.76~M_{\odot}$ in a 10-year survey would require an increase in number of observations per star from that in \citet{EPRVWGreport}.

\end{abstract}

\section{Introduction}\label{sec:introduction}
Future flagship mission concepts such as LUVOIR \citep{LUVOIR} and HabEx \citep{HabEx} will be capable of performing high-contrast direct imaging to characterize the atmospheres of nearby Earth analogs around Sun-like stars, and to search for potential biosignatures.  
Given the large observing time necessary for such detections, the efficiency of these missions benefits from a predefined target list of promising, potentially habitable near-Earth-mass candidates, rather than relying on their own blind direct imaging searches to supply these atmospheric characterization targets \citep{Dressing2019}. 
Transit surveys, e.g., \emph{Kepler} \citep{Borucki2010}, \emph{TESS} \citep{Ricker2014}, CHEOPS \citep{CHEOPS}, PLATO \citep{PLATO}, while capable of probing the habitable zones of low mass stars \citep[e.g.,][]{Anglada-Escude2016,Gillon2017,Gilbert2020}, will be unable to provide complete candidate target lists as their sensitivity to orbital period remains a challenge for the long orbital periods that correspond to the habitable zones of Sun-like stars. 
More importantly, the transit probability of such planets is $\approx0.5$\%, so we would expect to detect fewer than one transiting Earth analog among the $\sim$150 Sun-like stars within 15~pc \citep{Gupta2021}. 
In contrast, radial velocity (RV) surveys are sensitive to planets both at longer periods and at wider ranges of orbital inclinations and are therefore well suited to providing more complete candidate target lists.
Even if the results of an EPRV survey were not available in time to provide a target list for a direct imaging mission, measuring the planet masses with EPRVs would still be critical for interpreting direct imaging observations, e.g., providing an atmospheric scale height to interpret atmospheric properties, characterizing the extent of dynamical interactions with any other planets, and informing the planet's formation history \citep[e.g.,][]{vonParis2013,Nayak2017,Dorn2018,Suzuki2018}.

With this idea in mind, the NASA/NSF Extreme Precision Radial Velocity (EPRV) Working Group was established to study the feasibility of a large-scale RV survey to detect and precisely measure the masses of Earth analogs in support of future direct imaging missions, such as LUVOIR or HabEx \citep{EPRVWGreport}. 
One focus of the working group was to explore the specific design of such a survey, including telescope size, longitudinal coverage, and instrument capabilities, decisions which affect not only the science results of the mission but also the required cost to build and operate. 
To fully explore the impact of these decisions on the scientific goals to measure precise masses, \citet{EPRVWGreport} investigated ten ``architectures"---defined to encompass all decisions about telescope size, observing plan, instrument, longitudinal coverage, etc.---and simulated a 10-year survey of $\sim$100 nearby stars that are likely targets for future direct imaging missions. 
The promising results of these simulations found that multiple architectures could measure the masses of habitable zone Earth analogs to better than 10\%, under optimistic assumptions for the mitigation of stellar variability. 
A mass precision of 20\% or better is expected to result in uncertainties in atmospheric properties being dominated by the quality of spectroscopy and/or atmospheric modeling, rather than the mass uncertainty \citep{Batalha2019}.  
We follow the EPRV Working Group and adopt a more stringent target mass precision of 10\%, which is the necessary precision for distinguishing between composition models \citep[e.g.,][]{Valencia2007}, and also allows a margin for additional uncertainty due to stellar variability under more realistic assumptions. 
Despite being a conservative cutoff, achieving a mass precision of 10\% has additional benefits. 
For small eccentricities, the RV signal attributable to the orbital eccentricity is well approximated by the epicyclic approximation, i.e., a sinusoidal signal with amplitude $eK$.  Therefore, a mass precision of 10\% also implies an eccentricity precision of $\simeq$~0.1, which places stringent constraints on an orbital parameter that has implications for the formation and evolution of the Earth-like planet and its planetary system.

The survey simulations carried out for the EPRV working group simulation \citep{EPRVWGreport} were made under optimistic assumptions, assuming that each RV measurement has uncertainty dominated by photon noise and/or instrumental precision and that the measurement noise for each observation is independent and uncorrelated. 
The most straightforward interpretation of the results is that it assesses how RV surveys would perform if surveying stars with no intrinsic stellar variability (and any instrumental noise is uncorrelated).
An alternative more plausible interpretation is that it assesses how RV surveys would perform if improvements in the analysis of EPRV observations were to effectively remove all signals due to stellar variability (and any correlated instrumental noise).
Indeed, the spectral resolution, signal-to-noise and wavelength coverage for the survey architectures considered by the EPRV working group were chosen to provide information that could be used to mitigate stellar variability, based on analyses of simulated solar data \citep[e.g.,][]{Davis2017}.
At the moment, stellar variability---driven by many different astrophysical mechanisms, each with different amplitudes and timescales---remains the largest hurdle in the push toward detecting true Earth analogs in RVs \citep{Crass2021}. 
In order to properly characterize the efficiency of these simulated survey architectures, it is crucial to account for the stellar signals that will be imprinted in the radial velocities. 
In this study, we quantify the effects of stellar variability on the expected mass precision of the future surveys considered by the EPRV working group.

\subsection{Stellar Variability}\label{sec:stellarvariability}

\subsubsection{Magnetic Activity}
The longest timescale variations are due to stellar magnetic activity cycles, similar to the Sun's 11-year magnetic cycle, with times of high activity usually associated with higher RV variability. 
These periods of high activity are expected to correspond to an increase of surface activity features (faculae/plages, spots) that introduce variations on the star's rotational timescale, of order days to months. 
On the Sun, RV variability is dominated by faculae in large, concentrated regions of plages that locally suppress the convective blueshift \citep{Meunier2010,Haywood2016,Milbourne2019}.
Observed correlations between activity metrics (such as $\logRHK$, a measure of the chromospheric emission in the \ion{Ca}{2} H\&K lines) and radial velocity measurements \citep[e.g.,][]{Campbell1988,Saar1998,Santos2000,Wright2005,Isaacson2010,Luhn2020a} have led to some success identifying activity-induced signals using periodogram analyses \citep[e.g.,][]{Hatzes2010} and more recently using Gaussian process models that jointly fit simultaneous RV and activity time series \citep{Rajpaul2015,Gilbertson2020}. 
In some cases, additional information about rotationally modulated activity signals can be gleaned from high-precision photometry and can contribute to disentangling activity and planetary signals in RV time series \citep[e.g.,][]{Queloz2009,Aigrain2012,Haywood2014}.

However, \citet{Luhn2020a} find that while the RV RMS is positively correlated with such stellar activity metrics, activity-induced signals do not always manifest similarly in the RVs, with most stars showing positive correlation between individual RV measurements and simultaneous activity (indicative of local suppression of the ``net convective blue shift"), other stars showing no correlation (perhaps due to a pole-on inclination) and still others showing a negative correlation, perhaps indicating the stellar surface is dominated by bright magnetic features \citep[e.g.,][]{Milbourne2019}. 
Indeed, slowly rotating Sun-like stars have been shown to increase in brightness as their activity increases \citep{Radick1983,Lockwood1984,Lockwood2007} and so we expect these stars to have RV variability dominated by bright active regions.

\subsubsection{Granulation}
Stellar granulation, the surface manifestation of convection, operates on shorter timescales (minutes to hours).
Large, bright cells of hot rising gas cover the stellar surface, dominating over the darker, cool intergranular lanes of infalling material. 
This gives rise to the ``net convective blueshift" across the surface of a star \citep[e.g.,][]{Dravins1981,Gray2009,Meunier2010}. The convective cells vary in size (several hundred km to tens of Mm) and lifetimes (8 minutes to 1 day) from granulation to ``supergranulation" \citep{DelMoro2004,Hall2008}. 
In photometry, the granulation signal has been studied in both the time domain \citep[``8 hour flicker" seen in][]{Bastien2014} and in the frequency domain of asteroseismology \citep[e.g.,][]{Michel2008,Gilliland2010,Kallinger2014}. 
In RVs, \citet{CollierCameron2019} see evidence of granulation signals in the intraday variations of the daily high-cadence HARPS-N solar observations \citep{Dumusque2015}, matching the total granulation contribution expected ($\sim$0.5-1~\ms) for a solar-like star \citep{Meunier2015}. 
One strategy for mitigating the effects of granulation on RV surveys is to take multiple observations per night \citep{Dumusque2011a}.  
If the observations within a night are averaged while assuming each observation to be independent, then one would underestimate the uncertainty in the nightly RV.  
Therefore, it is important to consider the strength of correlation between observations when evaluating this or other EPRV observing strategies.
For a more in-depth review of the effect of granulation on the detection of Earth-mass planets, see \citet{Cegla2019}.

\subsubsection{Oscillations}
Stellar pressure-mode, or ``p-mode," oscillations operate on timescales of just minutes for solar-like stars.
The convective motions in a star generate acoustic pressure waves that ripple throughout the interior of star, creating deformations of the stellar surface. 
The deformations can be seen as intensity fluctuations in high-precision photometry of stars. 
\emph{Kepler} revealed a wealth of information about the general characteristics and behavior of p-mode oscillations of stars from asteroseismology analyses. 
In these analyses, the power spectral density (PSD) of a star reveals a bump of excess power due to oscillations, often modeled as a Gaussian, with characteristic frequency, amplitude, and width dependent upon stellar properties of the star, notably the surface gravity, $\logg$, and the effective temperature, $\Teff$ \citep[e.g.,][]{Brown1991,Kjeldsen1995,Kallinger2014}. 
In RV measurements, the oscillation bumps are more pronounced since the intensity fluctuations also correspond to material physically moving, compounding the effect. 
On the Sun, p-mode oscillations operate with a dominant timescale $\sim$5~minutes and amplitude $19$~\cms \citep{Chaplin2019}. 

RV surveys can effectively mitigate the effect of p-mode oscillations by choosing integration times that average over the dominant oscillation timescale \citep{Chaplin2019,Gupta2021}.
The EPRV working group survey simulations incorporated this observing strategy by setting a minimum exposure time of 5 minutes for each observation\footnote{When observing bright stars with large apertures, the survey simulations assumed multiple exposures spanning 5 minutes would be taken in rapid succession and binned into one observation.} and so we expect oscillations to be the least impactful source of correlated noise.

\subsection{Paper Organization}
In this work, we build on the results of \citet{EPRVWGreport} by including a more realistic model for correlated noise due to stellar activity, granulation, and oscillation in the calculation of planet mass precision. 
We choose Gaussian process (GP) noise models for each of the three key mechanisms responsible for stellar variability, following the prescription of activity in \citet{Gilbertson2020} and the model for oscillations and granulation in \citet{Pereira2019}. 
For the RV Community Challenge, \citet{Dumusque2016} took a slightly different approach, modeling oscillations and granulation components based on the power spectral density of these signals and adding an activity component based on simulations using SOAP~2.0 \citep{Dumusque2014}.
\citet{Hall2018} applied a similar SOAP~2.0-based activity component (but no granulation or oscillations) to simulated survey designs for the Terra Hunting Experiment. 
An advantage of the noise models in this work is that we scale granulations and oscillations based on stellar properties and use the GP formulation of these signals to construct the covariance matrix for a simulated set of observations, rather than assuming independent observations.

In \autoref{sec:architectures} we summarize the simulated EPRV surveys from \citet{EPRVWGreport}. 
In \autoref{sec:cov_kernels} we describe the GP covariance kernels used in this work, which are necessary for the calculation of the mass precision outlined in \autoref{sec:mass_prec}.
In \autoref{sec:analysis}, we analyze the effect of correlated noise on the simulated EPRV surveys and make comparisons to similar survey simulations with ideal and white noise models. 
Additionally, we investigate the choice of several survey parameters, including instrumental precision and survey duration. 
In \autoref{sec:discussion}, we discuss the implications of our results for planing EPRV surveys, consider caveats to the assumptions made in both the GP models and planet mass SNR calculation, and identify opportunities for further research. 
Finally, we summarize our conclusions in \autoref{sec:summary}.

\section{Simulated EPRV Planet Surveys}\label{sec:architectures}
We use the results of the simulated EPRV surveys in \citet{EPRVWGreport}, who ran a realistic observing program under various telescope architectures. 
Each architecture has distinct Northern and Southern Hemisphere surveys, assuming identical instruments, longitudinally distributed across existing observatories in each hemisphere. 
The simulations for each architecture include plausible estimates for the time allocation granted to the spectrograph on each telescope (varies with the telescope size) and take into account the observing season for each site and night-to-night observing window for each star. 
Additionally, the simulations account for typical weather losses. \autoref{tbl:defaults} gives a summary of the architecture default values.

For this study, we are concerned with the ability to detect Earth mass planets orbiting in the habitable zones of their stars and so we assume each star on the ``green target list'' has an Earth-mass habitable-zone planet in a circular orbit. 
In this case, the RV semi-amplitude scaling relation is
\begin{equation}
K = 9 \,\mathrm{cm \, s^{-1}} \left(\frac{m_p}{M_{\oplus}}\right) \left(\frac{P}{365 \, \mathrm{d}}\right)^{-\frac{1}{3}}  \left(\frac{M}{M_{\odot}}\right)^{-\frac{2}3} \sin{i},
\label{eqn:semi-amplitude}
\end{equation}
where the period of a habitable zone is determined by the mass of the star:
\begin{equation}
P = 365 \, \mathrm{d} \left(\frac{M}{M_{\odot}}\right)^{\frac{3\alpha -2}{4}},
\label{eqn:period}
\end{equation} 
assuming a MS luminosity scaling relation $L \propto M^{\alpha}$. 
We choose $\alpha=4$ for this work and assume orbits are viewed edge-on ($\sin{i} =1$).

\begin{deluxetable}{c c c c}
\tablecaption{Default parameters for simulated surveys. \label{tbl:defaults}}
\tablehead{
\colhead{Parameter} & Description & \colhead{Default value} & \colhead{Alternate values}}
\startdata
$T_{sur}$ & Survey Duration & 10~years & 2, 5, \& 15~years (\autoref{sec:survey_duration})\\
$\sigma_{inst.}$ & Instrumental precision & 5~\cms & 10 \& 30~\cms{} (\autoref{sec:instr})\\
N$_{targets}$ & Number of targets observed & 109 & --\\
\enddata
\end{deluxetable}

\begin{figure*}
\includegraphics[width=\textwidth]{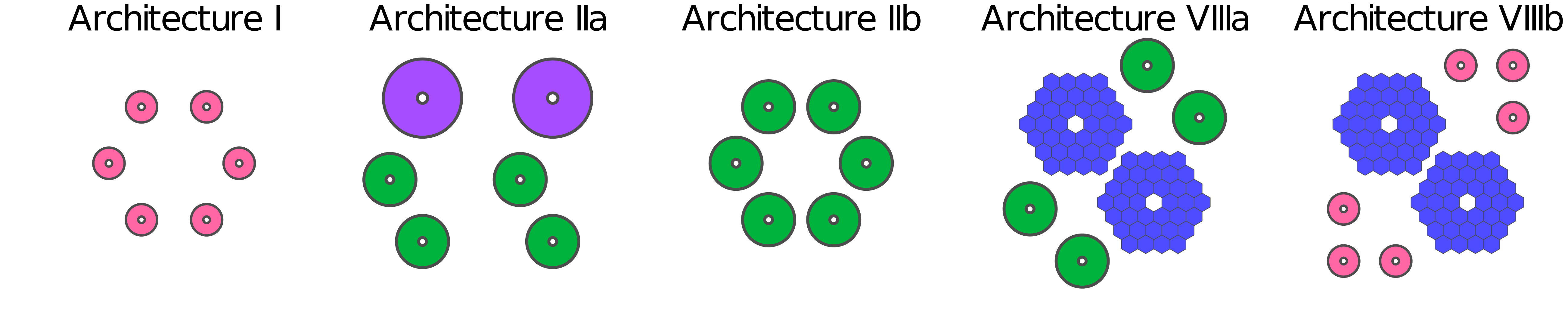}
\caption{Diagram summary of the 5 telescope architectures investigated in the study. Telescope diameters are to scale and are (from smallest to largest) 2.4~m (pink), 3.5~m (green), 6~m (purple), 10~m (blue, hexagonal). All architectures contain a Northern Hemisphere survey and a Southern Hemisphere survey (e.g., \I{} has 3 NH telescopes and 3 SH telescopes).}
\label{fig:architectures}
\end{figure*}

We focus on 5 in architectures in particular: Architectures \ToRoman{1}, \ToRoman{2}a, \ToRoman{2}b, \ToRoman{8}a, and \ToRoman{8}b. 
These were chosen to provide a variety of telescope number and sizes. 
\autoref{fig:architectures} shows a summary diagram of the telescopes that make up each architecture while \autoref{tbl:architecture_summary} shows a more detailed summary of the parameters for each telescope architecture.

Each survey observed the same 109 stars, which \citet{EPRVWGreport} refers to as the ``green target list''.
These stars had been previously identified as likely targets for one or more direct imaging mission concepts and are amenable to RV surveys based on their spectral type (F7-K9) and $v \sin i < 5$ km/s.
The observation time, exposure duration, and photon noise level of each observation in the simulations follow from the telescope aperture and properties of each star on the ``green target list''.

We summarize the details of each architecture below and refer the reader to \citet{EPRVWGreport} for a more in depth discussion of the characteristics of each architecture. 

\begin{deluxetable}{c c c l c c r r}
\tablecaption{Summary of Survey Architectures \label{tbl:architecture_summary}}
\tablehead{
\colhead{Architecture} & \colhead{$N_{obs,tot}$} &  \colhead{$\braket{N_{obs,star}}$} & \colhead{Telescopes} & \colhead{$N_{obs,tels}$} & \colhead{$\braket{N_{obs,tels,star}}$} & \colhead{$\braket{\sigma_{phot}}$} & \colhead{$\tau_{exp}$}}
\startdata
\tableline
\ToRoman{1}                            &                          308923  &                           2834 & \scriptsize{}6 $\times$ \small{}2.4~m & 308923 &  2834  & 9.2~\cms  & 8.4 - 31.7 min \\ \tableline
\multirow{2}{*}{\ToRoman{2}a} & \multirow{2}{*}{587813}  &  \multirow{2}{*}{5393} & \scriptsize{}2 $\times$ \small{}6.0~m & 285718 &  2621  & 7.0~\cms    & 4.1 - 6.6 min   \\
                                                 &                                        &                                    & \scriptsize{}4 $\times$ \small{}3.5~m & 302095 &  2772  & 9.0~\cms  & 5.3 - 14.4 min \\ \tableline
\ToRoman{2}b                          &                         515753   &                           4732 & \scriptsize{}6 $\times$ \small{}3.5~m & 515753 &  4732  & 8.9~\cms  & 5.3 - 14.0 min \\ \tableline
\multirow{2}{*}{\ToRoman{8}a} &  \multirow{2}{*}{437198} &  \multirow{2}{*}{4011} & \scriptsize{}2 $\times$ \small{}10~m  & 60354   &  554    & 4.0~\cms    & 3.3 - 9.0 min   \\ 
                                                 &                                        &                                    & \scriptsize{}4 $\times$ \small{}3.5~m & 376844 &  3457  & 11.5~\cms  & 5.1 - 8.2 min   \\ \tableline
\multirow{2}{*}{\ToRoman{8}b} &  \multirow{2}{*}{516225} &  \multirow{2}{*}{4736} & \scriptsize{}2 $\times$ \small{}10~m  & 60354   &  554    & 4.0~\cms    & 3.3 - 9.0 min   \\
                                                 &                                        &                                    & \scriptsize{}6 $\times$ \small{}2.4~m & 455871 &  4182  & 12.0~\cms  & 6.5 - 16.5 min \\
\enddata
\tablecomments{$N_{obs,tot}$ gives the total number of observations for a given architecture.
$\braket{N_{obs,star}}$ reports the mean number of observations per star ($N_{obs,tot}/109)$. 
The column labeled ``Telescopes" gives the telescope class breakdown for each architecture. For architectures made up of more than one telescope class, the following columns give the total number of observations in each telescope class $N_{obs,tels}$ and the subsequent mean number of observations per star in each telescope group $\braket{N_{obs,tels,star}}$. 
The final two columns give the mean photon precision, $\sigma_{phot}$, for an observation in that telescope class and the 25th and 75 percentile range for the exposure times, $\tau_{exp}$, for that telescope class.}
\end{deluxetable}

\subsection{Architecture I} 
\I{} contains six fully robotic 2.4~m telescopes \citep[based on the Automated Planet Finder,][]{Vogt2014} with Northern Hemisphere (NH) telescopes at Kitt Peak (KP), Calar Alto (CA), and Mauna Kea (MK) observatories and Southern Hemisphere (SH) telescopes at Las Campanas (LC), Sutherland (SA), and Siding Spring (SS) observatories. 
This architecture assumes each site is a dedicated facility for an EPRV survey and allocates 100\% of time for observing. 
Over the simulated 10-year survey, \I{} obtains 308923 observations (average 2834 per star). Note that \I{} in \citet{EPRVWGreport} used an instrumental uncertainty of 10~\cms. To better compare the effects of instrumental uncertainty in the presence of correlated noise, our default value for the instrumental uncertainty in \I{} is 5~\cms{}, to match the other architectures. We will later investigate in \autoref{sec:instr} how each architecture performs if we assume instrumental precision of 10 and 30~\cms.

\subsection{Architecture IIa}
Similar to \I{}, \IIa{} contains larger telescope designs to achieve higher precision for a fixed exposure. 
Each hemisphere is equipped with two 3.5~m telescopes (KP, CA in the north; SS, SA in the south) as well as a 6~m telescope (MK in the north; LC in the south). 
This architecture also assumes 100\% telescope time allotted to this survey. 
Over the simulated 10-year survey, \IIa{} obtains a total of  587813 observations (average 5392 per star) divided into 285718 observations across the two 6~m telescopes (2621 per star) and 302095 observations across the four 3.5~m telescopes (average 2772 per star).

\subsection{Architecture IIb}
\IIb{} is the same as \IIa{} but with the two 6~m telescopes replaced with 3.5~m telescopes, such that all six sites are identical.
Over the 10-year survey, \IIb{} obtains a total of 515753 observations (average 4732 per star). 

\subsection{Architecture VIIIa}
\VIIIa{} contains two 10~m telescopes at MK in the north and LC in the south, with the same four additional 3.5~m telescopes as in \IIa{} located at KP and CA in the north and SA and SS in the south. 
This architecture assumes 20\% telescope time allotted to the two 10~m telescopes (one night per week) and 100\% time on the four 3.5~m telescopes. 
Over the simulated 10-year survey, \VIIIa{} obtains a total of 437198 observations (average 4010 per star) divided into 60354 observations across the two 10~m telescopes (544 per star) and 376844 observations across the four 3.5~m telescopes (average 3457 per star).

\subsection{Architecture VIIIb}
\VIIIb{} is the same as \VIIIa{} but with six 2.4~m telescopes instead of four 3.5~m telescopes. 
As a result, there is a 10~m telescope and a 2.4~m telescope at both MK in the north and LC in the south. 
The two 10~m telescopes assume the same 20\% observing allocation as in \VIIIa{} (and the same observations). 
Over the simulated 10-year survey, \VIIIb{} obtains a total of 516225 observations (average 4736 per star) with 45871 observations across the six 2.4~m telescopes (average 4182 per star).

\section{Correlated Noise Models}\label{sec:cov_kernels}
\citet{EPRVWGreport} used survey simulations to establish the feasibility of measuring precise masses assuming that all stellar variability components could be removed. 
Under these optimistic assumptions, several architectures performed well, and would be able to measure planet masses with SNR $\gg$ 10 for most of the stars on the green target list. 
Given this result, choosing an architecture for the EPRV precursor survey could be largely driven by cost and feasibility. 
The goal of this study is to update these results with correlated noise models to investigate which architectures are well suited for detecting Earth-mass planets in the habitable zones in the presence of more realistic stellar variability models. 
We model the RV signal due to stellar variability as a Gaussian Process. 
A covariance kernel $k(\left|t-t'\right|)$, specifies the form of covariances as a function of the time lag between observations. 
We model the RV perturbations due to stellar variability as a sum of four terms, each a Gaussian Process: one term for activity, one term for oscillations, and two terms for granulation. 
We describe each in the sections below.


\subsection{Active Regions}\label{sec:spots_kernel}
We model RV perturbations due to rotationally linked stellar variability, hereafter ``Active Regions," or ``AR,'' using a GP model from \citet{Gilbertson2020}, hereafter \citetalias{Gilbertson2020}.
\citet{Gilbertson2020a} simulated time-series of solar spectra focusing on the effects of faculae/plages and spots (collectively ``active regions'') using a custom version of SOAP 2.0 \citep{Dumusque2014}.   \citetalias{Gilbertson2020} measured the apparent radial velocity and stellar variability indicators from each spectra and fit several Gaussian Process (GP) models to the resulting timeseries of stellar activity-induced RVs.  
Based on their results,  we adopt a linear sum of a Matern-5/2 GP kernel and its first derivative as the GP kernel for modeling the effects of active regions. 
Equation 9 of \citetalias{Gilbertson2020} shows the generalized form of the kernel. 
In the case of only using the RVs (i.e., not including additional time series such as activity indicators), the resulting GP kernel is simply a difference of the 0th order and 2nd order time derivatives of the kernel for a latent process (in this case the Matern-5/2 kernel), so we can write

\begin{equation}
k_{AR}\left(\left|t-t'\right|\right) = a_0^2 k_{M52}\left(\left|t-t'\right|\right) - a_1^2 \frac{d^2}{dt^2}k_{M52}\left(\left|t-t'\right|\right)
\label{eqn:spots_kernel}
\end{equation}

where $\{a_0, a_1\}$ are amplitude parameters (in m/s) and $k_{M52}\left(\left|t-t'\right|\right)$ is the Matern-5/2 kernel: 

\begin{equation}
k_{M52}\left(\left|t-t'\right|\right) = e^{\frac{-\sqrt{5}\left|t-t'\right|}{\lambda}} \left(1+\frac{\sqrt{5}\left|t-t'\right|}{\lambda} + \frac{5\left|t-t'\right|^2}{3\lambda^2} \right)
\label{eqn:m52}
\end{equation}
with timescale $\lambda$. We compute the second derivative of the kernel to derive the term in the right hand of \autoref{eqn:spots_kernel}  
\begin{equation}
\frac{d^2}{dt^2}k_{M52}\left(\Delta \right) = -\frac{R^2}{3} e^{-R\Delta}\left(1+R\Delta - R^2\Delta^2\right)
\label{eqn:m52_2nd}
\end{equation}
where we have used the shorthand $R \equiv \sqrt{5}/\lambda$ and $\Delta \equiv \left|t-t'\right|$. 

\begin{figure}
\centering
\includegraphics[width=0.6\columnwidth]{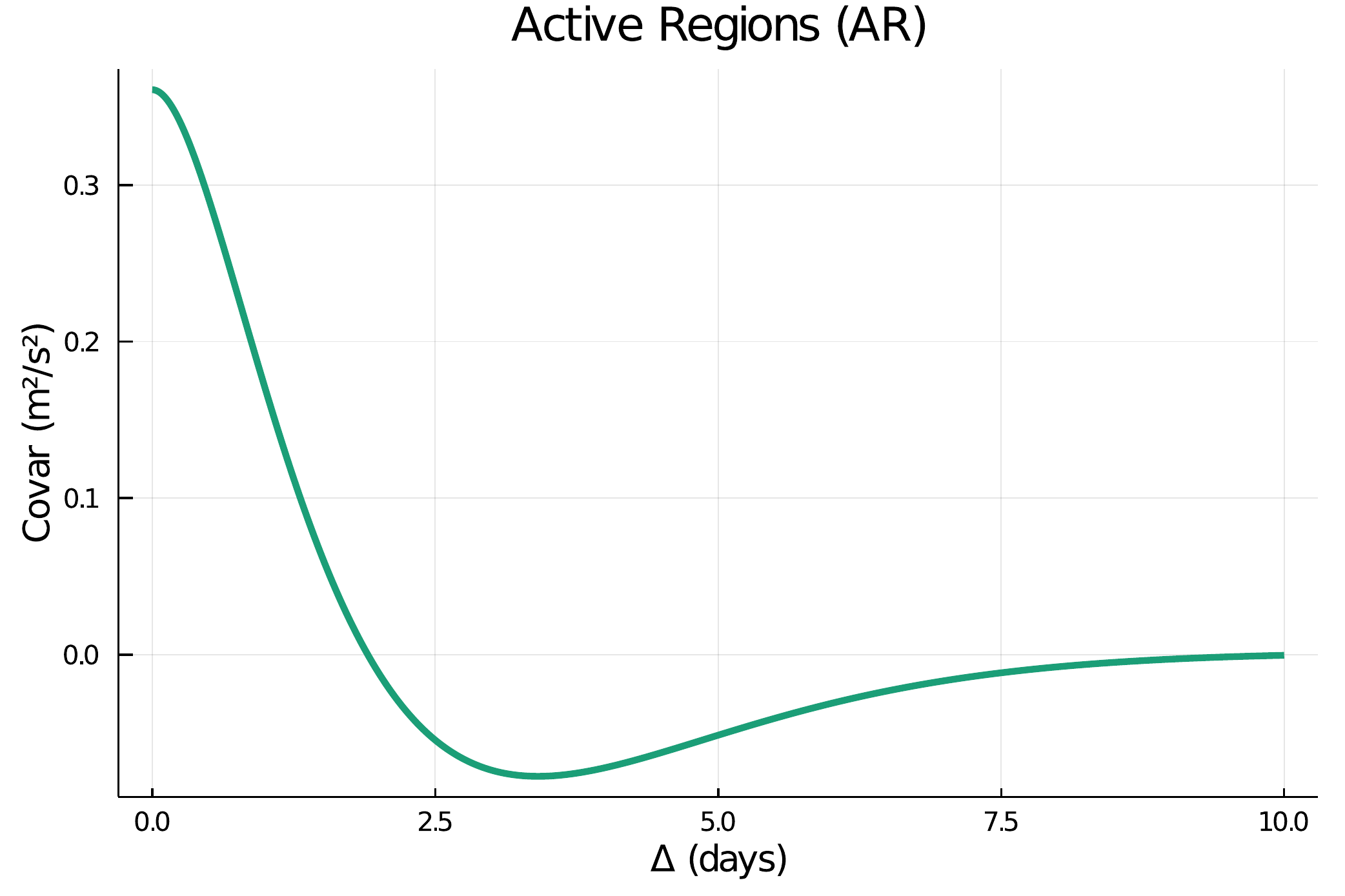}
\caption{The covariance kernel for the model of active regions (\autoref{eqn:spots_kernel}), which comes from \citetalias{Gilbertson2020}, a sum of a Matern-5/2 GP (timescale $\lambda = 2.52$~d, amplitude $a_0 = -0.106884~$\ms) and its first derivative (with amplitude $a_1=-1.154$~\ms). The covariance kernel is described by the absolute time difference between observations, $\Delta \equiv \left|t-t'\right|$. Note that these values have been fit to simulations of solar active regions; we use this same kernel for every star in our sample (i.e., we do not scale amplitude or timescale with stellar properties).}
\label{fig:spots_kernel}
\end{figure}

From solar simulations of \citetalias{Gilbertson2020}, the best fit for the solar timescale hyperparameter $\lambda$ is $2.52~$d, and the best-fit solar amplitudes are $\{a_0, a_1\} = \{-0.106884, -1.154\}$~\ms. The resulting covariance kernel is shown in \autoref{fig:spots_kernel}. 
The covariance matrix therefore has diagonal term ($ \left|t-t'\right|=0$) of 0.361 m$^2\,$s$^{-2}$, which implies a white noise term for active regions: $\sigma_{AR} = 0.6$ m$\,$s$^{-1}$. 
The correlated structure of activity-induced variability has not yet been well quantified as a function of spectral type. 
We adopt the solar values for all stars in our sample, unlike the oscillation and granulation kernels in the following sections, which we scale with stellar properties.
We discuss the potential impact of a model that depends on stellar parameters in \autoref{sec:disc:activity_dependence}.

\subsection{Stellar Oscillations} 
We model the RV perturbations due to stellar oscillations using a GP model from \citet{Pereira2019}, hereafter \citetalias{Pereira2019}.
\citetalias{Pereira2019} describe the transformation between the GP kernels for a stochastically driven, damped, harmonic oscillator---implemented in the \emph{celerite} framework described in \citet{Foreman-Mackey2017}---and the Gaussian functional form commonly used to approximate the stellar p-mode oscillation bump in asteroseismology power spectral density analyses. 
In the limit where the oscillator quality factor $Q\gg1$, the kernel becomes 

\begin{equation}
k_{osc}\left(\Delta\right) = S_{osc} \, \omega_{osc} \,Q \,e^{\frac{-\omega_{osc} \Delta}{2Q}}\left(\cos{\left(\eta \omega_{osc} \Delta\right)} + \frac{1}{2\eta Q} \sin{\left(\eta \omega_{osc} \Delta\right)} \right)
\label{eqn:covar_osc_inst}
\end{equation}

where $\Delta$ is the the absolute value of the time between observations as above, 
$\omega_{osc}$ is the characteristic frequency of the undamped oscillator, $S_{osc}$ is proportional to the power of the spectrum at $\omega = \omega_{osc}$, $Q$ is the quality factor, and $\eta = |1-(4Q^2)^{-1}|^{1/2}$ . \citetalias{Pereira2019} further give the relation between these kernel parameters and the asteroseismic parameters

\begin{equation}
\begin{split}
S_{osc} &= \frac{P_{g}}{4Q^2} \\
\omega_{osc} &= 2 \pi \nu_{max}
\end{split}
\end{equation}

where $P_g$ is the power excess of the oscillation bump and $\nu_{max}$ is the frequency of maximum p-mode oscillation power. 
When simulating stellar oscillations for stars other than the Sun, we adopt a scaling relationship for $\nu_{max}$ from \citet{Chaplin2019}

\begin{equation}
\nu_{max} = 3100\, \mu Hz \left(\frac{g}{g_{\odot}}\right)^{1} \left(\frac{\Teff}{T_{\mathrm{eff},\odot}}\right)^{-0.5}.
\end{equation}

To scale the power excess $P_g$ with stellar parameters, we follow \citet{Guo2022}, hereafter \citetalias{Guo2022}, who transformed photometric scaling relations into scaling relations for the apparent RV signal due to stellar variability, 

\begin{equation}
\begin{split}
P_{g, phot} (ppm^2   \mu Hz^{-1}) &= \sqrt{\frac{1}{2 \pi \sigma_{env}^2}} \left(\frac{a_{gran}}{4.57}\right)^{\frac{2}{0.855}} \\
r_{osc} &= 20 \beta\left(\Teff\right) \left(\frac{\Teff}{5777 \mathrm{K}}\right)^{-\alpha} \\
P_{g,RV} (m^2 s^{-2}  \mu Hz^{-1}) & = \frac{P_{g, phot}}{ r_{osc}^2}
\end{split}
\end{equation}

where $a_{gran} = 3382\nu_{max}^{-0.609}$ is the granulation amplitude\footnote{Note that this is the amplitude in photometry $(ppm)$. Refer to \citetalias{Guo2022} for a more complete breakdown.} \citep[][\citetalias{Guo2022}]{Kallinger2014}, $\sigma_{env} (\mu Hz) = 0.174 \nu_{max}^{0.88}$ is the width of the Gaussian power excess, $r_{osc}$ is the amplitude ratio between the photometry in the \emph{Kepler} bandpass and radial velocity power spectra, $\beta(\Teff)$ is the correction factor from bolometric flux to the \emph{Kepler} bandpass given by Equation 13 in \citetalias{Guo2022}, and $\alpha = 0.9$ \citepalias{Guo2022}. 
Therefore, given stellar parameters $\logg$ and $\Teff$, one can compute $\nu_{max}$ and from there calculate the kernel parameters in \autoref{eqn:covar_osc_inst}.

The GP kernel above (\autoref{eqn:covar_osc_inst}) describes the covariance between two instantaneous points in time. 
Unlike the active regions kernel in \autoref{sec:spots_kernel}, the timescale in \autoref{eqn:covar_osc_inst} is not always much longer than exposure times. 
Therefore, we compute the double integral for two observations with durations $\delta_1$ and $\delta_2$ and with observation start times separated by time $\Delta$. 
The integration can be performed analytically by parts. 
See \autoref{sec:osc_derivation} for the equations and derivation of the full oscillation covariance kernel and treatment of overlapping observations.

\begin{figure}
\centering
\includegraphics[width=0.6\columnwidth]{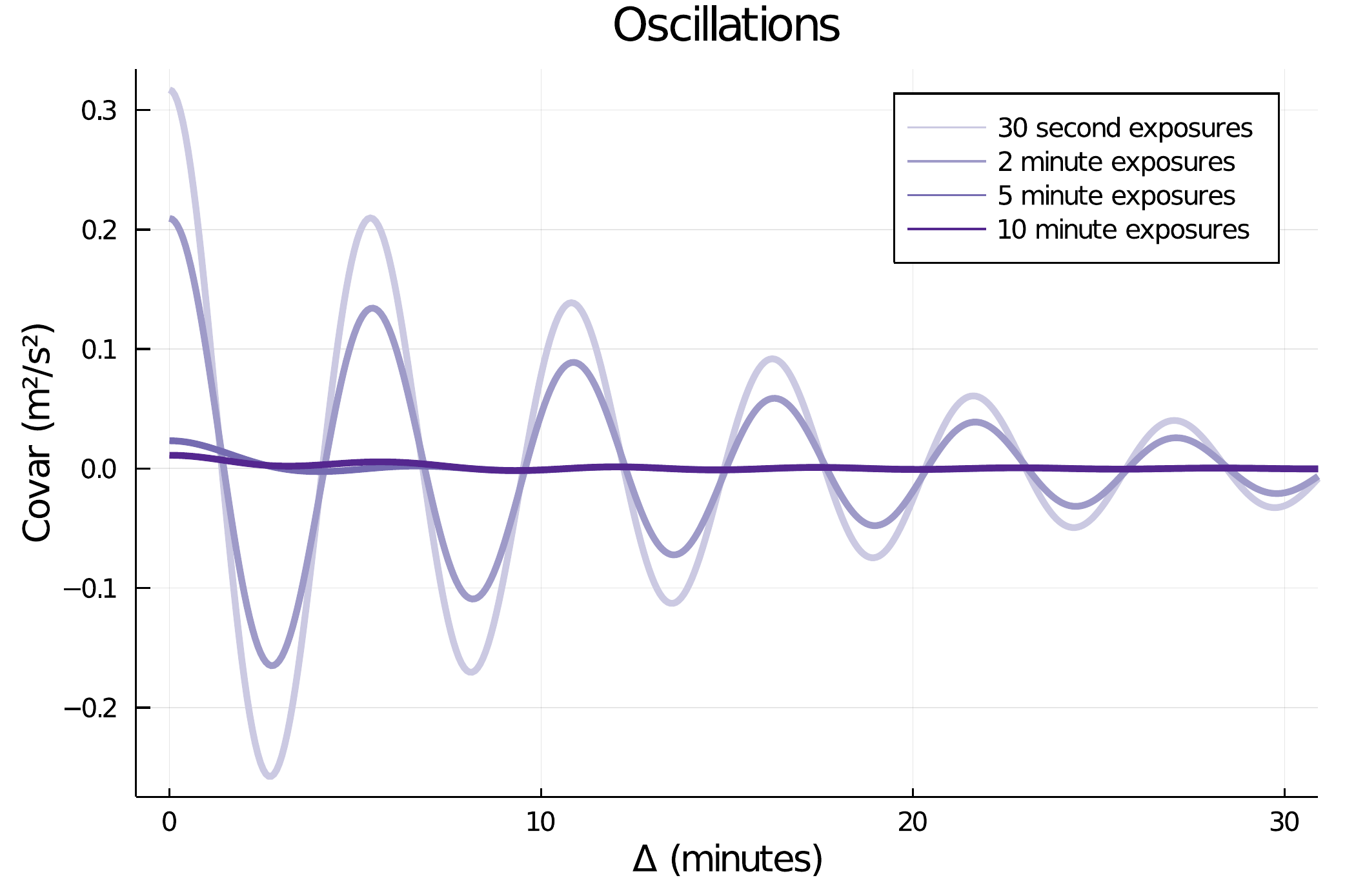}
\caption{The covariance kernel for the oscillation model for the Sun (Equations \ref{eqn:osc_sep} and \ref{eqn:osc_over}, with $\logg=4.43$, $\Teff =5777$~K). $\Delta$ is the time in minutes between a pair of observations and the exposure duration
is shown in different shades of purple. In this case, the two observations are the same duration ($\delta_1 =\delta_2$), as is typically the case for consecutive observations of a given star from the same telescope.}
\label{fig:osc_kernel}
\end{figure}


\subsection{Granulation}
We model the apparent RV signal due to granulation as a GP with two terms as in \citetalias{Guo2022},
\begin{equation}
\begin{split}
k_{gran}\left(\Delta\right) = S_1 \omega_1 e^{\frac{-\omega_1 \Delta}{\sqrt{2}}}\cos{\left(\frac{\omega_1 \Delta}{\sqrt{2}} - \frac{\pi}{4} \right)}  + S_2 \omega_2 e^{\frac{-\omega_2 \Delta}{\sqrt{2}}}\cos{\left(\frac{\omega_2 \Delta}{\sqrt{2}} - \frac{\pi}{4} \right)}
\end{split}
\label{eqn:gran_covar}
\end{equation}
where $S_1,  S_2$ define the power scaling and $\omega_1, \omega_2$ are the characteristic frequencies for each granulation component. 
When modeling granulation for stars other than the Sun, we adopt scaling relationships for $\omega$ from \citet{Kallinger2014} and \citetalias{Guo2022},
\begin{equation}
\begin{split}
\omega_1 & = 2 \pi b_1 = \left(2 \pi\right) 0.317 \nu_{max}^{0.97} \\
\omega_2 & = 2 \pi b_2 = \left(2 \pi \right) 0.948 \nu_{max}^{0.992}.
\end{split}
\end{equation}
Similarly, $S_1$ and $S_2$ are calculated using photometric scaling relations to convert to RV scaling relations. First, following \citetalias{Pereira2019}, we adopt
\begin{equation}
S_{n} = \frac{a_{n}^2\sqrt{2}}{\omega_{n}}.
\end{equation}
We set $a_1 = a_2 = a_{gran}$ in RVs (as in photometry, \citetalias{Guo2022}) and rescale the amplitudes for the photometric signature of granulation
\begin{equation}
a_{gran,phot} (ppm) = 3382 \nu_{max}^{-0.609} \\
\end{equation}
to obtain the amplitude for the RV signature of granulation
\begin{equation}
a_{gran, RV} (m/s) = \frac{a_{gran, phot}}{r_{gran}}
\end{equation}
As with the oscillation kernel, the granulation kernel can be scaled to a given star using $\logg$ and $\Teff$. 

\begin{equation}
r_{gran} = 100 \left(\frac{\Teff}{5777 \mathrm{K}}\right)^{-32/9} \left(\frac{g}{g_{\odot}}\right)^{2/9} \\
\end{equation}

We follow the same approach as for the oscillation component for the granulation component to compute the double integral across two observations. See \autoref{sec:gran_derivation} for the equations and derivation of the full granulation covariance kernel and treatment of overlapping observations.

\begin{figure}
\centering
\includegraphics[width=0.6\columnwidth]{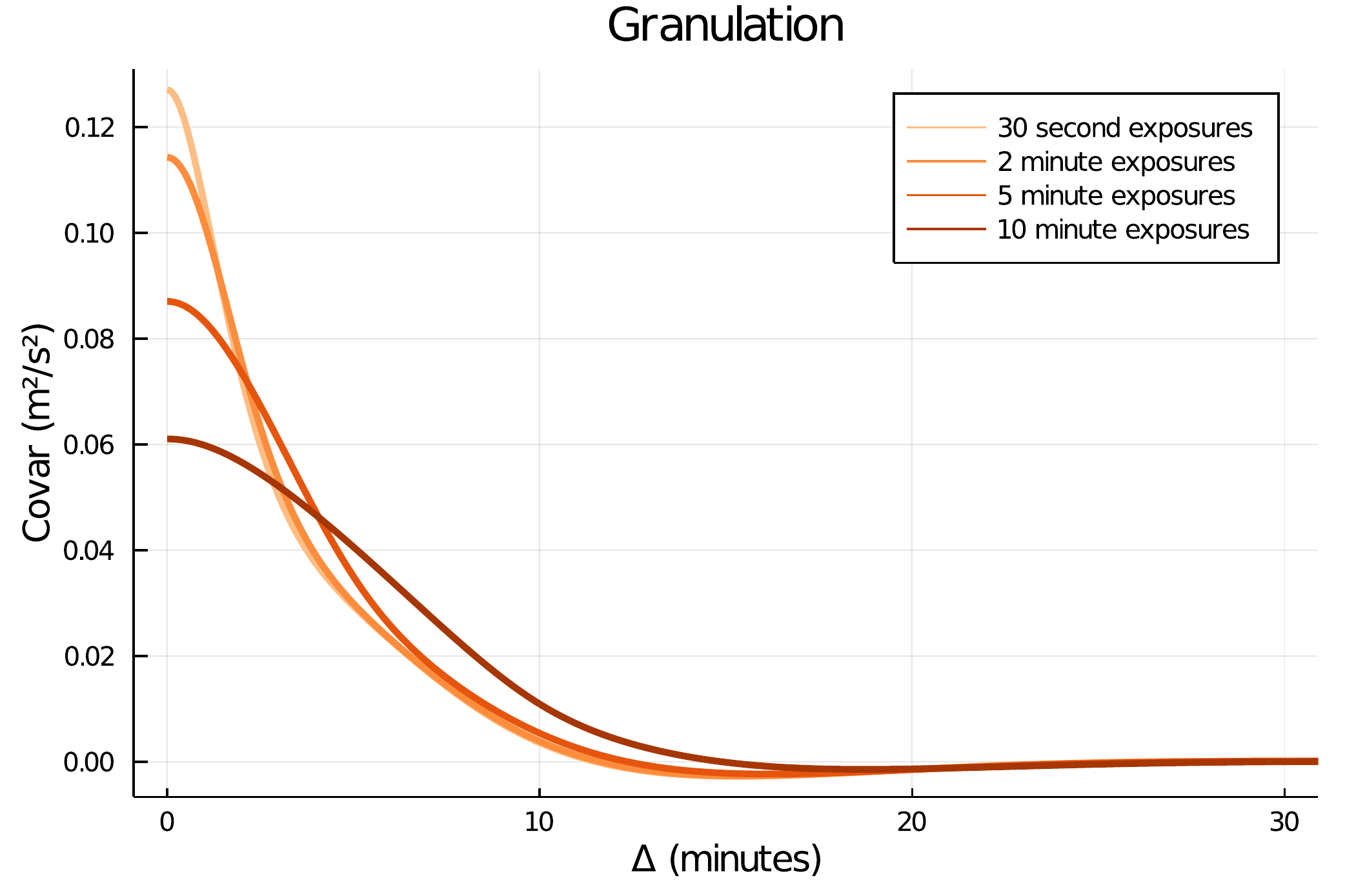}
\caption{The covariance kernel for the granulation model for the Sun (Equations \ref{eqn:gran_sep} and \ref{eqn:gran_over}, with $\logg=4.43$, $\Teff =5777$~K). $\Delta$ is the time in minutes between a pair of observations and the exposure duration
is shown in different shades of red-orange. In this case, the two observations are the same duration ($\delta_1 =\delta_2$), as is typically the case for observations of a given star from the same telescope.}
\label{fig:gran_kernel}
\end{figure}

\section{Estimating the Mass Precision}\label{sec:mass_prec}
We wish to estimate the measurement precision for $K$, the radial velocity semi-amplitude due to a putative planetary signal. 
After defining the covariance kernels for each of the activity components in \autoref{sec:cov_kernels}, we can compute the full covariance matrix $\bm{C}$ as a sum of the covariance kernels described in \autoref{sec:cov_kernels} for each pair of observations $i,j$ and adding diagonal terms due to photon and instrumental noise:
\begin{equation}
C_{ij} = k_{AR, ij} + k_{osc, ij} + k_{gran, ij} + \delta_{ij} \sigma^2_{phot,i} + \delta_{ij} \sigma^2_{instr}
\label{eqn:covar_matrix}
\end{equation}
where $\delta_{ij}$ is the Kronecker delta and $\sigma_{phot,i}$ is the photon noise for observation $i$, and $\sigma_{instr}$ is the constant instrumental uncertainty term.

From this we can calculate the uncertainty in measuring the semi-amplitude, $K$, assuming planets are on circular orbits such that
$$\bm{v} = K \sin{\left(\frac{2\pi\bm{t}}{P} + \phi_0\right)}$$ 
where $\bm{v}$ are the true, kinematic radial velocities at observation times $\bm{t}$, for a planet with orbital period $P$ (given in \autoref{eqn:period}) and phase $\phi_0$. If we assume the orbital period and phase are well known, then generalized least squares gives the parameter uncertainty in the estimate of $K$
\begin{equation}
\sigma^2_{K} =  \left(\bm{X}' \bm{C}^{-1} \bm{X}\right)^{-1}
\end{equation}
where
\begin{equation}
\bm{X}  = \sin{\left(\frac{\bm{t}}{P} + \phi_0\right)}.
\label{eqn:design_matrix}
\end{equation}
Often, the orbital period and phase are well known when performing RV follow-up of a planet identified via transit follow-up surveys.  
While direct imaging could provide secure planet detections, the uncertainty of the orbital period and phase will be greater than is typical of planets discovered via transit.
When the orbital period is unknown, fitting for all parameters at once makes the problem non-linear, so generalized least squares is no longer appropriate.  
Unfortunately, performing a Markov chain Monte Carlo analysis \citep[e.g.,][]{Ford2006} is orders of magnitude more computationally demanding than generalized least squares. 
The formal uncertainty from generalized least squares provides a more direct comparison to the results of \citet{EPRVWGreport} and represents both an approximation to and a lower bound on the uncertainty if one were to allow for potential correlations with orbital period and phase.
We justify the assumption of well known periods through an investigation of simulated RV time series with 1000 realizations of planet phase and stellar variability---drawn from the GP kernels described previously in \autoref{sec:cov_kernels}---and find that the simulated RV time series results in a significant periodogram peak\footnote{0.1\% FAP} at the ``correct'' period (within 3\% of the true planet period) at a median success rate of 96\% across all stars in this sample \footnote{\citet{Nava2020} performed a similar analysis using a quasiperiodic rotation kernel, which is on longer timescales than the timescales used in our AR model. However, while their results show strong peaks near the rotation signals and its harmonics, they still rarely observed peaks near the long periods of interest in this work ($\sim$365 days for a true Earth anolog). Further, the simulated RVs in \citet{Nava2020} spanned only a few observing seasons, with $\sim$100 total observations. The signals of the planets in our simulated periodograms are strengthened significantly (and the AR signals made less coherent) by the 30-50 times more observations over a baseline 10 times as long for the simulated surveys studied in this work.}. 
We discuss implications of this assumption in \autoref{sec:disc:periods}. 
We also wish to emphasize that while the focus of this study is the precision of the mass SNR for Earth analogs, it will be important to obtain planet mass measurements that are not only precise, but also accurate.

For the purposes of this study, we are primarily interested in understanding the impact of correlated noise due to stellar variability on the yield of EPRV surveys for potentially Earth-like planets.  
Therefore, we compare the results of various surveys, using generalized least squares to approximate the uncertainty in each planet mass for each survey.
We calculate the signal to noise ratio for the planet semi-amplitude estimate, $\hat{K}$, 
\begin{equation}
\mathrm{SNR} = \frac{K}{ \sigma_K},
\end{equation}
which corresponds to the SNR for the estimated planet mass.
Note that this SNR quantifies the precision of the mass estimate and is not necessarily a useful criterion for detecting a planet, nor does it provide information about the nature of the signal. 
Traditionally, astronomers have refrained from claiming a detection of an exoplanet until the data show that the orbital period is well constrained and coherent.  
Adopting such a criterion would require exploring a large parameter space (e.g., computing a periodogram) and checking whether the power near the true period is significantly larger than both a detection threshold and the power at any other putative orbital period.  
Such caution is particularly important when attempting to discover planets that may have short orbital periods and/or when using a relatively small number of observations, due to potential aliasing of the true period with the observational window function.
Aliasing will be less important for the EPRV surveys explored in \citet{EPRVWGreport} and this study, since these focus on orbital periods near a year and each star is to be observed thousands of times.  
Nevertheless, quantifying the significance of a planet detection is complicated and computationally demanding once one allows for the possibility of multiple planets and non-circular (or even non-Keplerian) orbits \citep{Nelson2020}.  
Further, direct imaging of a planetary system could provide independent detection of planets and constraints on their orbits.  
For the purposes of both \citet{EPRVWGreport} and this study, we use generalized least squares to approximate the uncertainty for the mass estimate, assuming that the orbital period and phase are well known.
This is in line with other assumptions (e.g., circular orbit, any other planetary signals are removed perfectly, etc.) that mean the uncertainties we compute will underestimate the uncertainties of a fully realistic EPRV survey.

Because we have fixed the planet sizes and periods and rely on the simulated observing times from \citet{EPRVWGreport}, the only remaining parameter that can affect the final $SNR$ is the phase $\phi_0$, i.e., how the observations line up with the planet's orbital phase. To account for this, we draw 20 random samples of $\phi_0$ between 0 and 2$\pi$ and report the median SNR for each planet and survey architecture.

In summary, we perform simulations asserting that each star has an Earth-mass planet in a circular, edge-on orbit in its habitable zone. 
Rather than performing a signal injection and recovery analysis, we instead use the covariance matrix following generalized least squares to calculate the SNR for the resulting planet's Doppler amplitude. 
This allows for quick, efficient computation of the expected performance for each telescope architecture. 

We compare the resulting uncertainties under 3 analysis models for the ``noise'' due to stellar variability: correlated stellar noise, white noise (independent and identically distributed, or iid), and no stellar noise (ideal star). 

\begin{itemize}
\item{\textbf{Correlated noise (corr)}: The correlated noise model calculates the covariance matrix in \autoref{eqn:covar_matrix} including all off-diagonal elements between observation pairs.}
\item{\textbf{White noise (iid)}: The white noise model uses only the diagonal terms of the covariance matrix, such that stellar variability is assumed to only contribute independent white noise (``jitter'') to each observation. This is equivalent to adding a Kronecker delta term in front of each stellar variability kernel in \autoref{eqn:covar_matrix}, such that only $k(\Delta =0)$ is evaluated for each mechanism.} 
\item{\textbf{No stellar variability (ideal)}: The no stellar variability model only includes the photon noise and instrumental uncertainty as diagonal elements in the covariance matrix (i.e., each stellar variability component kernel in \autoref{eqn:covar_matrix} is 0). This model is equivalent to the analysis in \citet{EPRVWGreport}. }
\end{itemize}

\section{Analysis}\label{sec:analysis}

\subsection{Architecture Performance for Baseline Survey for Ideal, White, and Correlated Noise Models}\label{sec:all_analysis}
\begin{figure}
\includegraphics[width=\columnwidth]{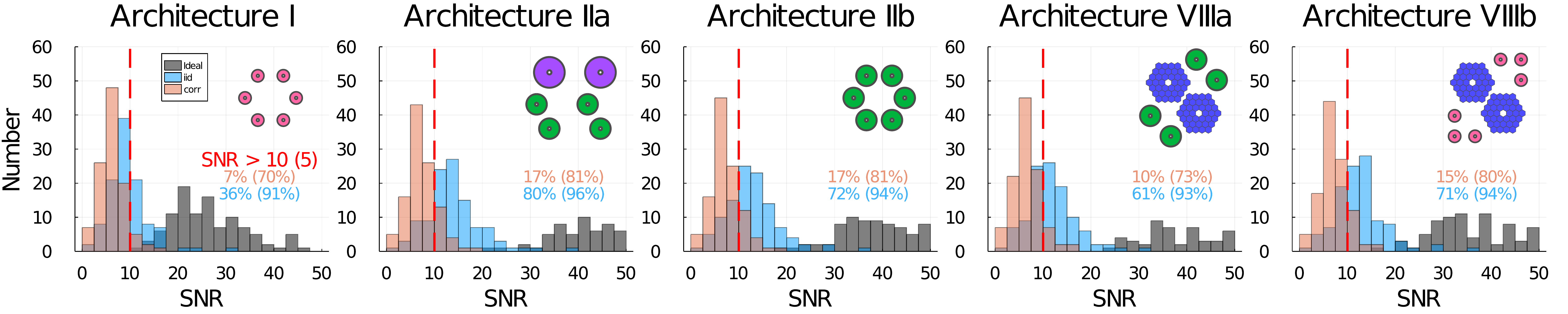}
\caption{Histograms of RV $SNR$ for measurements of Earth-mass planets for proposed RV survey architectures and various noise models for the effects of stellar variability on RV measurements. 
Each column shows a different telescope architecture, as described in \autoref{sec:architectures} and \autoref{fig:architectures}. 
Diagrams for each telescope architecture are repeated in the top right of each histogram for easy reference. 
Histograms are colored according to the assumptions for how stellar variability affects the RV measurements: correlated noise (``corr" in red/pink), white noise (labeled as ``iid") in blue, and no additional noise due to stellar variability (``ideal") plotted for just the top row in black/grey. 
The vertical dashed red line indicates where planet mass measurements will have a precision of 10\% or better.
The exact percentage of planets above this threshold is given in red and blue in the bottom right, corresponding to the correlated and white noise models, respectively. 
We also include in parentheses the percentage of planets with mass SNR~$>$~5 (20\% mass precision).
These data were generated using the default survey duration of 10 years and instrumental precision $5$~\cms. Note that the percentage of planets with mass measurement SNR~$>$~10 in the ``ideal" scenario (shown in the black/grey histograms) is 100\% for every architecture, meaning that the labels also indicate the reduction expected by including the specified noise source.}
\label{fig:noise_model_comparison}
\end{figure}

In this section we perform the analysis described in \autoref{sec:mass_prec} to explore the effect of each of the various correlated noise models on the precision of mass measurements.
We compute the set of SNR's for each star's mass measurement using the full noise model (active regions, granulation, and oscillations) assuming the nominal 10-year survey with 5~cm\,s$^{-1}$ instrumental precision. 
The results are shown in \autoref{fig:noise_model_comparison}, where we summarize the performance of each architecture by reporting the number of planets measured with mass SNR $>10$ (i.e., 10\% planet mass uncertainty),  followed by the number with SNR $>5$ (i.e., 20\% planet mass uncertainty) in parentheses. 
We also report in \autoref{tbl:architecture_performance} the median SNR (along with the 25th and 75th percentiles) for each architecture.

While each of these architectures is capable of measuring Earth-analog planets with mass precision better than 10\% in the ``ideal'' case, we find that is not the case when accounting for additional noise due to stellar variability.
First, we look at the white noise (``iid'') models and see that \I{} easily performs the worst as the only architecture with less than 50\% of planets measured with SNR $ > 10$. We expect this from \autoref{tbl:architecture_summary}, since \I{} has the fewest number of observations per star.  
The relative performance of the rest of the architectures can be predicted based on the average number of observations per star.
This follows expectations that in a white noise scenario, performance between architectures will depend largely on the number of observations.

In simulations assuming correlated noise due to stellar variability, the number of planets with masses measured with SNR~$>$~10 drops drastically relative to either white noise or ideal simulations. 
From the white noise to the correlated noise case, we expect to measure planets masses to 10\% or better for 4-6 times fewer planets, depending on the architecture.
At best, we expect only 17\% of planet masses to be measured to better than 10\%.
That is, \emph{in the presence of correlated noise due to stellar variability, a 10-year survey with $\sim$5000 observations per star and instrumental precision of 5~\cms{} is not likely to measure planet masses to better than 10\% for the majority of habitable Earth-mass planets around Sun-like stars}, given the assumptions of our correlated noise models (e.g., solar-like activity amplitudes) and the underlying architecture simulations (e.g., assuming no coordination between telescopes and no further progress in the ability to mitigate the effects of stellar variability beyond the high cadence and large number of observations).
Despite this, the histograms show that the majority of planet masses are measured with $5 \leq$SNR$\leq 10$ in all architectures, with the worst architecture still capable of measuring planet masses to better than 20\% for 70\% of stars.
With a SNR~$>$~5 threshold, is is expected that the mass precision (20\% or better) will lead to atmospheric characterization with uncertainties dominated by spectroscopic uncertainties and atmospheric modeling \citet{Batalha2019}.
While each architecture is able to achieve this threshold for a majority of the targets on the ``green target list," there is a steep drop off as the threshold is increased.
Due to the steep drop off, the number of stars with SNR~$>$~5 or SNR~$>$~10 will be sensitive to the specifics of our correlated noise model for stellar variability.
Similar to \citet{EPRVWGreport}, we default to discussing results in terms of performance under a SNR~$>$~10 threshold. 
However, where \citet{EPRVWGreport} used this conservative threshold as a way of accounting for stellar variability, we quantitatively consider the survey performance under a plausible correlated noise model.  Nevertheless, one could consider  adopting the same conservative threshold (SNR~$>$~10) to allow for \emph{additional} variability effects that are not captured in our model (see discussion in \autoref{sec:discussion}).

We can quantitatively describe the reduction in the SNR from the ideal to white noise to correlated noise models. 
We define the median ratio of the SNR from one noise model to another using the notation
\begin{equation}
\widetilde{R}^{ideal/iid} \equiv Med\left(\frac{\mathrm{\bf{SNR}}_{ideal}}{\mathrm{\bf{SNR}}_{iid}}\right)
\label{eqn:snr_ratio}
\end{equation}

We report the values of $\widetilde{R}^{ideal/iid}$, $\widetilde{R}^{iid/corr}$, and $\widetilde{R}^{ideal/corr}$ in \autoref{tbl:architecture_performance}. Across the five architectures, the white noise model reduces the SNR by a factor of $\sim$3-6 from the ideal model, and the correlated noise model reduces the SNR by a factor of $\sim$2 from the white noise model (a factor of $\sim$4-9 from white noise).
\begin{deluxetable}{c c c c c c c}
\tablecaption{Architecture Performance Under Ideal, White, and Correlated Noise Models \label{tbl:architecture_performance}}
\tablehead{
\colhead{Architecture} & \colhead{SNR$_{ideal}$} &  \colhead{SNR$_{iid}$} & \colhead{SNR$_{corr}$} & \colhead{$\widetilde{R}^{ideal/iid}$}  & \colhead{$\widetilde{R}^{iid/corr}$} & \colhead{$\widetilde{R}^{ideal/corr}$}} 
\startdata
\ToRoman{1} & 26.0$_{20.6}^{32.1}$ & 9.1$_{7.3}^{11.8}$ & 6.0$_{4.9}^{7.5}$ & 2.81 & 1.51 & 4.21 \\
\ToRoman{2}a & 51.3$_{41.0}^{66.4}$ & 13.0$_{10.6}^{16.0}$ & 7.1$_{5.4}^{8.8}$ & 3.98 & 1.84 & 7.42 \\
\ToRoman{2}b & 43.3$_{36.2}^{55.2}$ & 12.2$_{9.8}^{15.4}$& 7.0$_{5.4}^{8.7}$ & 3.62 & 1.73 & 6.30 \\
\ToRoman{8}a & 54.4$_{38.0}^{73.0}$ & 11.4$_{9.0}^{13.8}$ & 6.4$_{4.8}^{8.1}$ & 5.63 & 1.76 & 9.47 \\
\ToRoman{8}b & 39.9$_{31.4}^{53.3}$ & 12.2$_{9.5}^{14.8}$ & 7.0$_{5.3}^{8.6}$ & 3.40 & 1.74 & 6.03 \\
\tableline
\enddata
\tablecomments{Columns labeled SNR report the median SNR value for the given architecture and noise model, with the 25th and 75th percentile values given in subscript and superscript, respectively. The columns labeled $\widetilde{R}$ are calculated according to \autoref{eqn:snr_ratio}.}
\end{deluxetable}

It is also useful to compare the performance between similar architecture pairs (\IIa{} vs. \IIb{} and \VIIIa{} vs. \VIIIb{}) in the correlated noise case to see how the architecture design choice is affected by correlated noise. Comparing \IIa{} and \IIb{}, we see they perform similarly, despite \IIa{} having, on average, 660 more observations per star (a 14\% increase). The telescope number and location are the same between the two architectures, and the increased number of observations comes from the shorter exposure times on the two larger 6~m telescopes in \IIa{}. These observations also have photon precision of 7~\cms{} instead of the 9~\cms{} photon precision for the 3.5~telescopes. Despite these benefits for \IIa{}, the improvements are not distinguishable. Comparing \VIIIa{} and \VIIIb{}, we do see improved performance in \VIIIb{} over \VIIIa{}. Despite the lower photon noise from the 2.4~m telescopes, the additional telescope site more than makes up for it due to increased number of observations (on average 18\% more observations per star) and longitudinal coverage. Remember that in \VIIIa{} the 10~m telescopes only observe a star once per week on average. In \VIIIb{}, the additional telescope in both the northern and southern hemispheres are located at the same observatory as the 10~m telescopes, and therefore this gap in cadence and longitudinal coverage is filled.

In the next sections we look at the performance of each architecture under each of the ideal, white and correlated noise models as we change 1) which astrophysical mechanisms are included in the noise models, 2) the survey duration, and 3) the instrumental uncertainty.

\subsection{Individual Effects of Active Regions, Granulation, and Oscillation}\label{sec:isolated_analysis}
\begin{figure}
\includegraphics[width=\columnwidth]{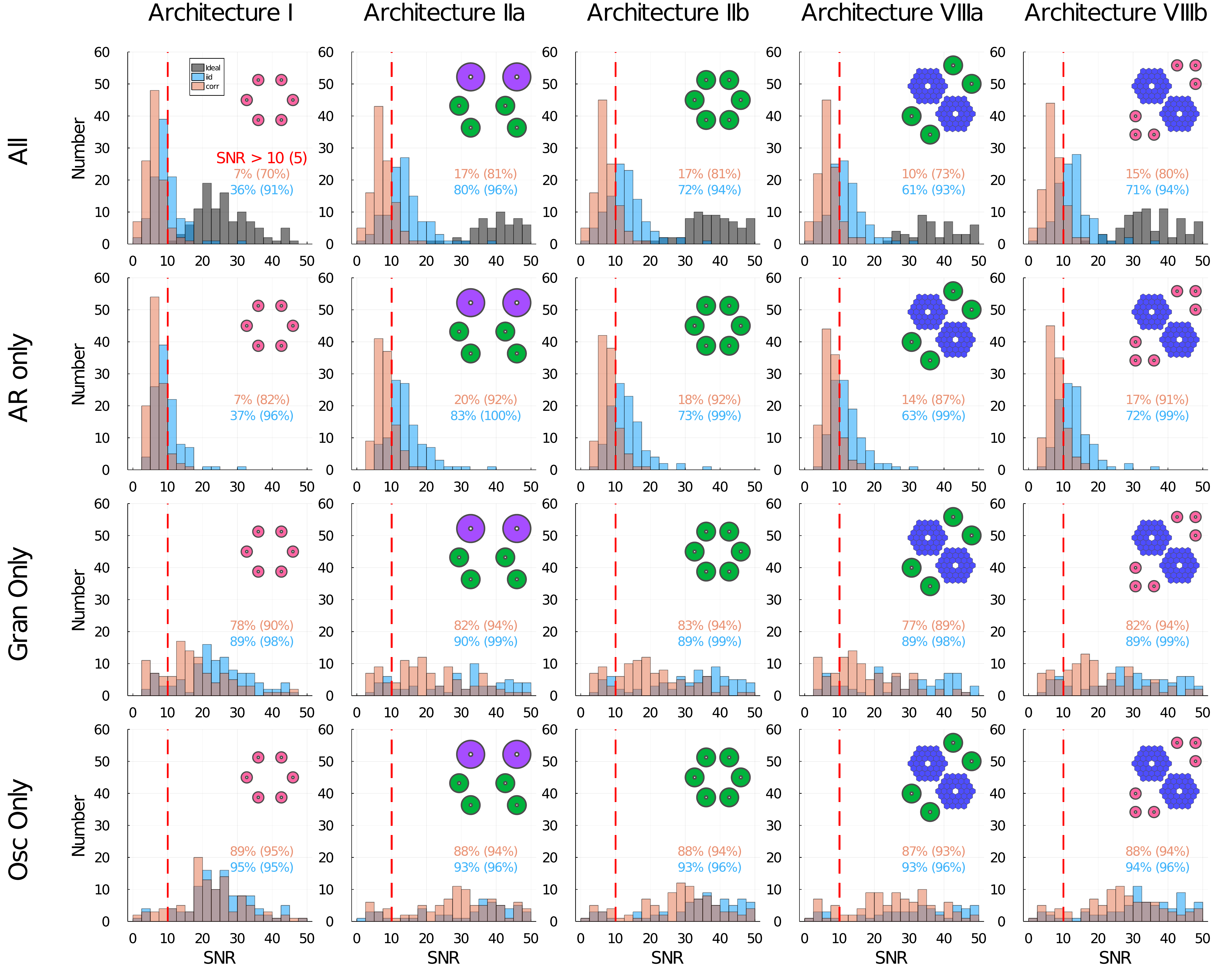}
\caption{ Same as \autoref{fig:noise_model_comparison} but with additional rows showing the results of simulations that include the effects of only one of three mechanisms (i.e., active regions, granulation, oscillations) for how stellar variability affects the observed RVs.
For each panel, we used the default survey duration of 10 years and instrumental precision 5~\cms. 
The black/grey ``ideal" histograms are only shown in the top row for easier legibility, since they are the same for each row.}
\label{fig:component_hist}
\end{figure}

Here we examine the effect of each mechanism individually. 
We follow the same procedure as above, calculating the formal errors assuming the data are generated and analyzed with matching noise models by including only one of the three stellar variability kernels at a time ($k_{AR}$, $k_{osc}$, and $k_{gran}$ in \autoref{eqn:covar_matrix}).
We note that we expect oscillations to be least impactful due to our choice of exposure times \citep[following][]{Chaplin2019}.
The results are shown in \autoref{fig:component_hist}. 

Let us first consider the impact of each noise mechanism on \IIa{} (see second column of Fig \autoref{fig:component_hist}). 
We can use the same ratio notation as above, but we now include subscripts to denote the stellar variability mechanism, e.g., $\widetilde{R}^{ideal/iid}_{osc}$ refers to the ratio of the mass SNR in the oscillation-only ideal model to the mass SNR in the oscillation-only white noise model.
If we consider only the white noise model (blue histograms) for each mechanism, then the impact is small for oscillations and granulation, which reduce the mass SNR from the no variability (``ideal") case by a median factor of $\widetilde{R}^{ideal/iid}_{osc}=1.01$ and $\widetilde{R}^{ideal/iid}_{gran}=1.14$, respectively. 
However, the impact is significant for active regions, which reduces the mass SNR from the ideal case by a median factor of $\widetilde{R}^{ideal/iid}_{AR}=3.96$.
If we consider the correlated noise models one noise source at a time, then each of these factors is increased and the reduction in the mass SNR is more severe. 
Active regions are still the most important effect, reducing the mass SNR by a median factor of $\widetilde{R}^{ideal/corr}_{AR}=6.64$, but the impact of granulation is also significant, reducing the mass SNR by a median factor of $\widetilde{R}^{ideal/corr}_{gran}=2.5$. 
As in the white noise case, the effect of oscillations is small, reducing the mass SNR from the ideal case by only a factor of $\widetilde{R}^{ideal/corr}_{osc}=1.51$.
If we compare the AR-only noise model to including all three noise sources, then we see that the combination is slightly worse than AR-only, presumably due to the increased effect of granulation.
In total, the combined effect of all variability sources reduces the mass SNR by a median factor of 7.42 in the correlated case, compared to a median factor of 3.98 in the white noise case. 
These values, along with the values for each architecture are reported in \autoref{tbl:isolated_performance}.
We conclude that correlated ``noise'' due to active regions and granulation is an important effect that must be considered when forecasting the science yield of an EPRV survey similar to those proposed by the \citet{EPRVWGreport}.
\begin{deluxetable}{c c c c c}
\tablecaption{Individual Effects of Active Regions, Granulation, and Oscillation on Mass SNR \label{tbl:isolated_performance}}
\tablehead{
\colhead{Architecture } & \colhead{Mechanism} & \colhead{$\widetilde{R}^{ideal/iid}$}  & \colhead{$\widetilde{R}^{iid/corr}$} & \colhead{$\widetilde{R}^{ideal/corr}$}} 
\startdata
 \multirow{4}{*}{\ToRoman{1}} & All & 2.81 & 1.51 & 4.21 \vspace{2pt} \\
& AR-only & 2.81 & 1.39 & 3.96 \\
& Gran.-only & 1.01 & 1.43 & 1.49 \\
& Osc.-only & 1.00 & 1.08 & 1.09 \\
\tableline
 \multirow{4}{*}{\ToRoman{2}a} & All & 3.98 & 1.84 & 7.42 \vspace{2pt} \\
& AR-only & 3.96 & 1.71 & 6.64 \\
& Gran.-only & 1.14 & 2.00 & 2.49 \\
& Osc.-only & 1.01 & 1.47 & 1.51 \\
\tableline
 \multirow{4}{*}{\ToRoman{2}b} & All & 3.62 & 1.73 & 6.30 \vspace{2pt} \\
& AR-only & 3.55 & 1.61 & 5.69 \\
& Gran.-only & 1.07 & 1.82 & 2.18 \\
& Osc.-only & 1.01 & 1.32 & 1.34 \\
\tableline
 \multirow{4}{*}{\ToRoman{8}a} & All & 5.63 & 1.76 & 9.47 \vspace{2pt} \\
& AR-only & 5.62 & 1.57 & 8.18 \\
& Gran.-only & 1.32 & 2.21 & 3.27 \\
& Osc.-only & 1.03 & 1.84 & 2.01 \\
\tableline
 \multirow{4}{*}{\ToRoman{8}b} & All & 3.40 & 1.74 & 6.03 \vspace{2pt} \\
& AR-only & 3.40 & 1.61 & 5.46 \\
& Gran.-only & 1.10 & 1.77 & 2.09 \\
& Osc.-only & 1.01 & 1.31 & 1.33 \\
\tableline
\enddata
\end{deluxetable}

Next, we compare the projected number of Earth-analog detections for the five most promising EPRV survey architectures considered by \citet{EPRVWGreport}, while accounting for the effects of correlated ``noise'' due to stellar variability.
Qualitatively, each of the survey architectures considered by \citet{EPRVWGreport} is affected similarly, but the magnitude of the effect differs due to the typical number and spacing of observations dictated by the survey architecture.
Considering all three sources of correlated noise, \IIa{} performs the best.
To understand the reasons, it is useful to consider the relative performance of each survey architecture in the presence of each type of stellar variability.

First, we consider the white noise scenario and note that the differences between architectures are minimal for granulation-only and oscillation-only.
\I{} performs the best in the oscillation-only scenario, as expected since this can be explained by the long exposure times on the 2.4~m telescopes in this architecture (refer to \autoref{tbl:architecture_summary}) that average over the typical p-mode oscillations more effectively than architectures with larger telescopes and shorter exposure times\footnote{For \I{}, the median exposure time was 2.8 times the characteristic oscillation timescale, compared to 1.3 times the characteristic oscillation timescale in \IIa{}. Following \citet{Chaplin2019}, this results in a residual oscillation amplitude in \I{} roughly 2 times lower than in \IIa{}.}.
Other architectures perform worse than \I{} in the oscillation-only scenario, despite having $\sim$40--90\% more observations per star, due to their shorter integration times. 
However, we emphasize that the performance difference between architectures in the oscillation-only case is minimal and not drastically different than the ideal model, suggesting that oscillations will not be the dominant hurdle to measuring precise masses of Earth-mass exoplanets in future surveys, provided they adopt similar exposure times.

Differences between architectures in the white noise granulation-only case are even smaller. For most stars, the granulation timescales are longer than integration times for any of the telescopes considered, so exposure time is not as effective in mitigating granulation as it is for oscillations. 
Instead, granulation mitigation occurs from observing a given star multiple times per night across the multiple telescopes \citep{Dumusque2011a,Meunier2015}.
\IIa{} performs only slightly better, likely due to the larger number of observations.

The biggest differences between architectures in the white noise model come from the AR-only scenario, with results closely matching those described in \autoref{sec:all_analysis}. 
Since the AR model has the largest white noise amplitude, the number of observations per star accounts for the performance differences between architectures. 
Note the poor performance of \I{} when accounting for the effects of active regions.
Even in the white noise scenario, less than 40\% of stars are detected with SNR~$>$~10, nearly a factor of two worse than the next best architecture, \VIIIa.

Finally, we compare the yield of different survey architectures in the presence of correlated ``noise'' from each type of stellar variability.  Qualitatively, the story is similar. 
For each mechanism, the SNR of mass measurements decreases relative to a white noise model.
Again, stellar variability due to active regions is the dominant mechanism for reducing the survey yield relative to ideal or white noise simulations.  
\IIa{}, \IIb{}, and \VIIIb{} perform the best in the presence of AR-induced noise, but the number of planets with masses measured at SNR~$>$~10 decreases significantly to at most 20\% of stars surveyed, even in the absence of granulation or oscillations. 
While either granulation or oscillations alone does not significantly reduce the number of planets with mass SNR~$>$~10, each does reduce the expected mass SNR significantly for most architectures (excluding \I{}; \IIb{} shows only a modest effect).
Despite the fact that neither granulation nor oscillations alone significantly reduces the number of planets with mass SNR~$>$~10, each does still introduce a reduction in the overall planet mass SNR in most architectures.
This additional reduction due to granulation and oscillations is what accounts for the reduction in the number of planets with masses measured at SNR~$>$~10 when comparing the simulations including all three mechanisms for stellar variability to the AR-only model.
In other words, AR-induced variability is the mechanism that reduces the planet mass precisions to the SNR~$\sim$~10 level; once at this level, it is important to account for the additional reduction due to granulation (smaller) and oscillations (smallest) to estimate the number of stars above the SNR~$>$~10 threshold.
These differences between architectures are reflected when using SNR~$>$~5 threshold as well, with improved overall performance across the architectures, as discussed in \autoref{sec:all_analysis}.

\subsection{Varying Survey Duration}\label{sec:survey_duration}
\begin{figure}
\includegraphics[width=\columnwidth]{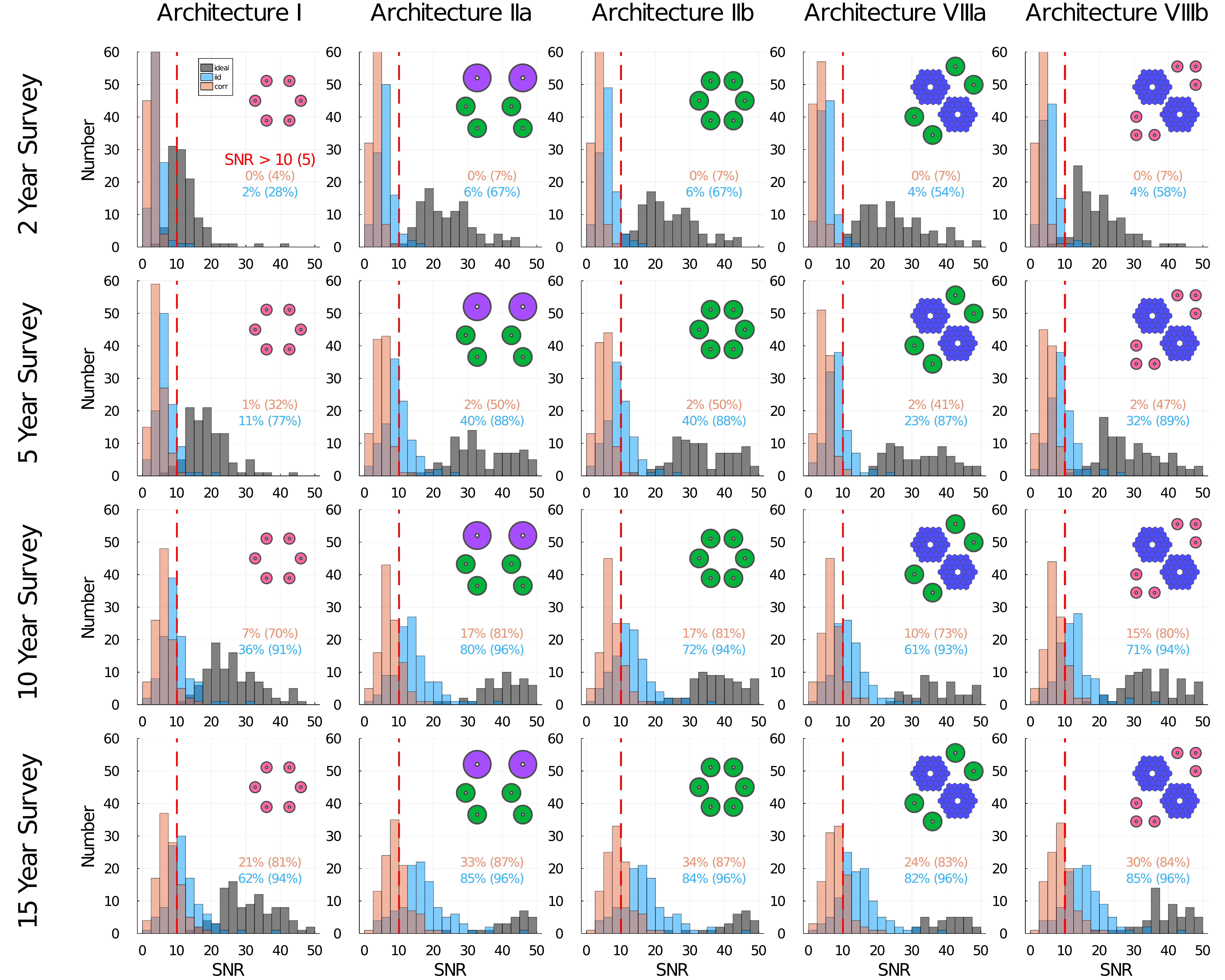}
\caption{Same as \autoref{fig:component_hist} but now showing the effect of survey duration on the distribution of mass SNR for each architecture. 
In this case, we have used the default 5~\cms{} instrumental noise and included all stellar noise components (active regions, granulation, oscillations) in both the correlated (corr) and white noise (iid) calculations. Note that the third row (10-year survey) is then identical to \autoref{fig:noise_model_comparison}.}
\label{fig:duration_hists}
\end{figure}

Next, we explore the effect of survey duration, considering 2-, 5-, 10-, and 15-year surveys. 
In the case of survey durations less than 10 years, we truncate the simulated observation time series, noting that this induces a proportional decrease in the number of observations during the course of the survey. 
Given the previous results, we expect poor performance from the 2 year and 5 year surveys; we include these as reference points for other instrument surveys \citep[e.g.,][]{Gupta2021}. 
To simulate a 15-year survey, we populate each additional year beyond year 10 with a year randomly selected from the previous 10 years. 
The results are shown in \autoref{fig:duration_hists}.

The SNR increases as the survey duration (and thus the number of total observations) increases.
An important result is the substantial benefit of a 15-year survey in terms of number of planet masses measured with SNR~$>10$.
In both correlated and white noise models, the SNR scales such that a 50\% increase in number of observations results in an increase in SNR by a factor of $\sim$1.22, as expected from $\sqrt{N}$, implying the improvements are due to the increased number of observations rather than increased baseline.
In almost every architecture, increasing the survey duration from 10 to 15 years (and thus increasing the planet mass SNRs by 22\%) results in twice as many planet masses measured with SNR~$>10$.

Note, however, that our AR model does not account for long-timescale variations in activity as seen in the activity cycles of many stars, including the Sun. 
While the added observations from extending to a 15-year survey will certainly improve the SNR of planet detections, the benefits may therefore be overestimated under this model.
We discuss the possibilities of accounting for and modeling long-term activity cycles in \autoref{sec:disc:cycles}.

\subsection{Varying Instrumental Uncertainty}\label{sec:instr}
\begin{figure}
\includegraphics[width=\columnwidth]{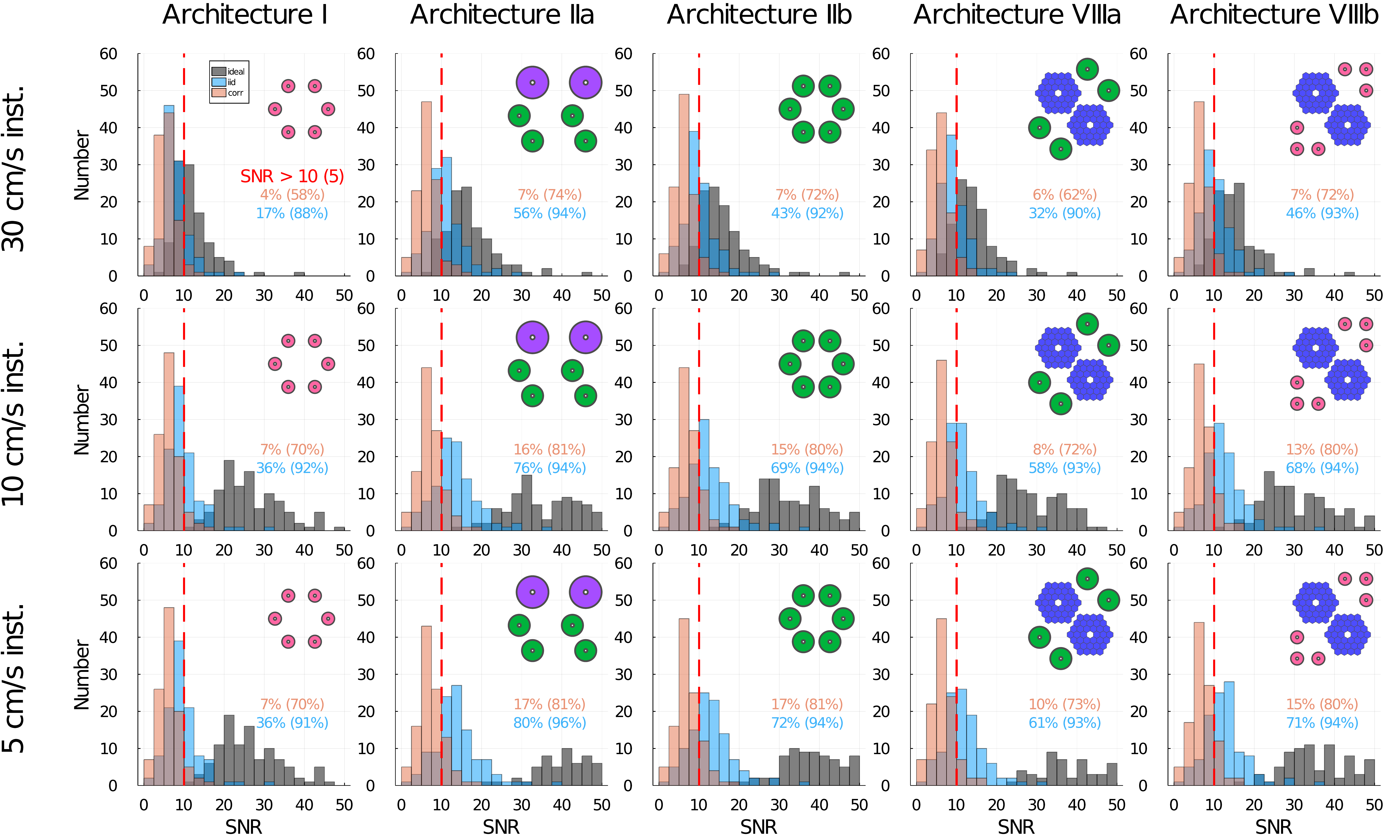}
\caption{Same as \autoref{fig:noise_model_comparison} but now showing the effect of instrumental precision on the overall distribution of SNR for each architecture (\autoref{fig:noise_model_comparison} appears in this figure as the bottom row). In this case, we have used the default 10-year survey duration and  included all stellar noise components (active regions, granulation, oscillations) in both the correlated (corr) and white noise (iid) calculations. As expected, the 5~\cms{} instrumental precision performs the best. However, there is not a significant improvement for most telescopes by going down to 5~\cms{} from 10~\cms. For many of the telescopes in these architectures, improving the instrumental precision from 10~\cms{} to 5~\cms{} means improvements down to the photon noise level. The slight improvements seen in \IIa{}, \VIIIa{}, and \VIIIb{} come from the 6~m and above telescope classes, which have photon noise below 10~\cms{}  (refer to \autoref{tbl:architecture_summary}).}
\label{fig:precision_hists}
\end{figure}

Finally, we explore the effect of instrumental precision ($\sigma_{instr}$) on survey yield. 
Our noise model (\autoref{eqn:covar_matrix}) assumes each observation is contaminated by 5~\cms, 10~\cms, and 30~\cms{} of non-calibratable instrumental noise that is Gaussian and uncorrelated between observations.
For this case, we return to the default 10-year survey and include all 3 astrophysical noise sources (active regions, granulation, and oscillations), both correlated and uncorrelated, as well as the ideal scenario with no stellar variability. 
The results are shown in \autoref{fig:precision_hists}.

Regardless of the survey architecture, there is a major increase in yield (for planets above SNR~$>$~10) as the instrumental precision improves from 30~\cms{} to 10~\cms{}.
However, improving the instrumental precision from 10~\cms{} to 5~\cms{} results in minimal improvement in survey performance. 
For many of the telescopes in these architectures, improving the instrumental precision from 10~\cms{} to 5~\cms{} means improvements down to the photon noise level (refer to \autoref{tbl:architecture_summary}).
Since many of the telescopes have photon noise level between 5 and 10~\cms, improving the instrumental precision below 10~\cms{} has diminishing returns given that the photon noise is added in quadrature with the other variability components.
The slight improvements seen in \IIa{}, \VIIIa{}, and \VIIIb{} come from the 6~m and above telescope classes, which have one telescope per hemisphere with photon noise of 7, 4, and 4~\cms{}, respectively, whereas \I{} and \IIb{} have photon noise $\sim$9~\cms.
This result is promising as current efforts are underway to improve from the state-of-the-art $\sim30$~\cms{}, with 5~\cms{} instrumental precision still considered an optimistic goal \citep[e.g.,][]{Pepe2010,Mahadevan2012,Schwab2016,Jurgenson2016,Szentgyorgyi2016}.

\subsection{Dependence on Stellar Mass}\label{sec:stellar_prop}
\begin{figure}
\includegraphics[width=\columnwidth]{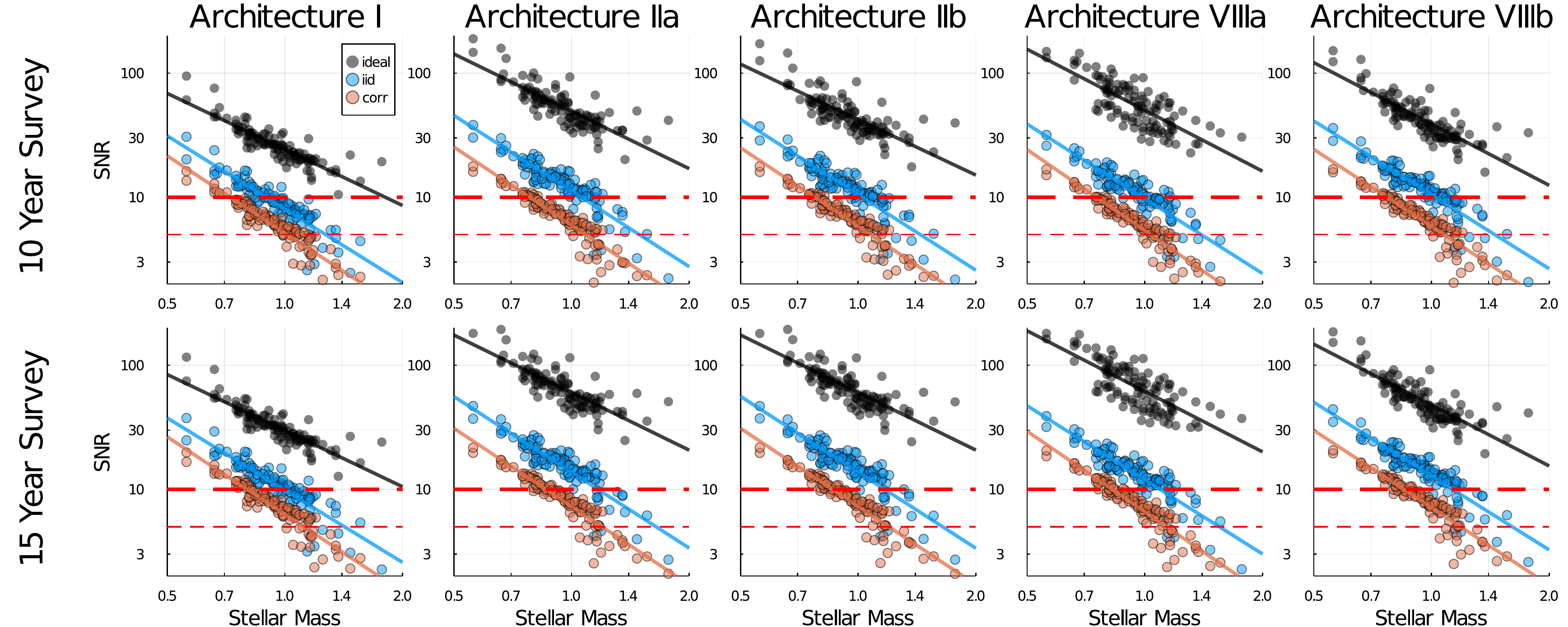}
\caption{Planet mass SNR as a function of stellar mass for each architecture, including no noise (``ideal"), white noise (``iid"), and correlated noise (``corr"). We show this for the default 10-year survey (top row) as well as the 15-year survey (bottom row), both using the default 5~\cms{} instrumental noise. Notice the strong trend with mass and the effect that introducing noise (correlated or uncorrelated) has on the overall SNR for these planets. In particular, in no case are we able to measure a planet mass with SNR above 10 (and therefore measure masses to 10\% precision) for a solar mass star. In fact, the cutoff mass above which we cannot achieve SNR$ = 10$ is closer to 0.76~$M_{\odot}$ (K2).}
\label{fig:mass_plots}
\end{figure}

Finally, we investigate the effect of host star mass on the SNR of planet mass measurements, shown in \autoref{fig:mass_plots}. Given the promising improvement by extending the survey duration,  we include results for both 10 and 15-year survey durations in \autoref{fig:mass_plots}.
Following \autoref{eqn:semi-amplitude} and \autoref{eqn:period}, we expect the signal of an Earth-mass planet in the habitable zone to be proportional to $M^{-1.5}$. 
However, fitting a power law to the mass SNR reveals power-law slopes closer to -2.
This indicates that $\sigma_K \propto M^{0.5}$ under our model of stellar variability (both white and correlated noise). 
Since our model for stellar variability due to active regions does not depend on the stellar properties, the scaling $\sigma_K \propto M^{0.5}$ must be due to effects of the combination of granulation and oscillation. 
It will be important for future studies to examine correlated noise models for magnetic activity to establish how the timescales and amplitudes vary with spectral type.
Including a scaling of active regions with stellar type would result in a different slope, as would a scenario in which the effectiveness of active region mitigation strategies depended on stellar mass.
Conversely, we may find that our current AR model uniformly over- or underestimates the amplitude of rotationally linked, activity-induced variability. 
Correcting for this would result in a uniform vertical offset for each star, with the overall slope unchanged.
We discuss the impact of an activity model dependent on stellar properties in \autoref{sec:disc:activity_dependence}.

\autoref{fig:mass_plots} shows which target stars would be expected to yield a mass measurement with SNR~$>$~10 (or 5) if they host an Earth-analog based on our survey simulations.
In the correlated noise case, most mass measurements have a SNR of 5-10. 
If one adopts a mass SNR threshold of 10 then the EPRV surveys considered would be limited to planets around low mass stars, $M_{\star}\lesssim0.76~M_{\odot}$ (K2) for a 10-year survey and $M_{\star}\lesssim0.85~M_{\odot}$ (K0) for a 15-year survey. 
While this paints a bleak picture for measuring precise masses of true Earth-analogs with present instruments and data analysis techniques, we note that each of the five architectures considered here is capable of measuring the mass of Earth-analogs with SNR~$>$~5 or better up to $M_{\star} \sim 1.1~M_{\odot}$. 
Obtaining more precise mass measurements is also possible if the field were to make significant improvements in our ability to mitigate stellar variability.

\section{Discussion}\label{sec:discussion}

\subsection{Comparison of Different Survey Architectures}
In our simulations, \IIa{} performs the best across all tests, with \IIb{} performing similarly well. 
While neither is capable of measuring more than 20\% of planet masses with SNR~$>$~10 (except in the case of a 15-year survey), they are both routinely capable of detecting more than 80\% of planet masses with SNR~$>$~5 in most scenarios, including in the case of 10~\cms{} instrumental precision.
Given the relatively poor performance of \VIIIa{} under the correlated noise models, our simulations suggest that similar RV surveys may want to disfavor larger (10~m) telescopes, if they indeed come at the cost of fewer observations, as prescribed in these architectures. 
However, there may be additional benefits of larger telescopes to correct or characterize stellar variability in detail, which is not considered in these simulations, e.g., validating MHD simulations of magnetoconvection \citep{Cegla2019}.
We note that \IIb{} outperformed \VIIIb{}, despite having roughly the same number of observations per star.  This result suggests that an array of small (2.4~m) telescopes with photon noise above 10~\cms{} is also disfavored.
Therefore, it seems that there is a sweet spot in terms of telescope size, coverage, and photon precision.
Our simulations favor architectures with a longitudinally spaced array of 3.5~m telescopes with 100\% observing time and 7-10~\cms{} instrumental precision.  
We recommend that the size and observing time of observatories for an EPRV survey be revisited once the community learns more about how effectively the effects of stellar variability can be mitigated as a function of  spectral resolution, SNR and wavelength coverage.

\subsection{Implication for planning EPRV surveys}
\subsubsection{Active Regions \& Rotationally Linked Variability}
The architecture that performed best was able to measure planet masses to better than 10\% for only 17\% of the sample in a 10-year survey with $\sim$5000 observations, and for only 33\% of the sample in a 15-year survey with $\sim$7500 observations. 
Given our results that rotationally linked variability was the dominant mechanism, we therefore find that a large number of observations is needed to overcome active regions, if using purely time-domain methods.
Given the larger number of observations required by the surveys simulated in this study, we encourage further research to develop stellar variability indicators that would enable future surveys to overcome the challenges of stellar variability with fewer observations.
Our results here highlight the importance of ongoing efforts on other fronts to account for and mitigate stellar variability \citep[see, e.g., Table A-3 of][]{Crass2021}.
Additional improvements (relative to our results) may be possible if combined with simultaneous activity time series, which have not been accounted for in these calculations.
Further improvements may be also possible if we are able to correct stellar activity in wavelength domain \citep[e.g.,][]{Davis2017,Dumusque2018,Wise2018,CollierCameron2020,deBeurs2020,Holzer2021,Cretignier2021}.
With these approaches, larger apertures may still be valuable or even critical, if our ability to correct for or characterize stellar variability depends on SNR and/or resolution.

\subsubsection{Granulation \& Oscillations}
Increasing telescope size has diminishing returns due to integration time and the combination of oscillations and granulation.
While smaller telescopes make fewer total number of observations for a given survey duration and target list, the longer exposure times on smaller telescopes helps them to suppress the amplitudes of spurious RV signals due to oscillation and granulation.

It is important to note that the stellar sample studied here contains mostly solar analogs, which allow for proper averaging over the dominant stellar p-mode timescale (5 minutes for the Sun).
Additionally, the granulation timescales are such that observations spanning multiple telescopes in a given night effectively average in a manner similar to \citet{Dumusque2011a}, who used multiple observations per night using a single telescope.
These observation strategies may not be as effective for surveys of more evolved stars, with longer timescales for granulation (days) and oscillation (hours).
Additional improvement for mitigating granulation may be possible following \citet{Dumusque2011a} on each telescope, rather than each architecture as a whole, though at the cost of a reduced target list in order to observe each star more frequently.
Since granulation is currently not the limiting factor in the success of these surveys, we deem such measures as non-critical.
As above with activity, advancements in mitigation techniques in the wavelength domain \citep{CollierCameron2020} may require new observational approaches to mitigate granulation that could favor larger apertures or higher cadence.

Considering only oscillations and granulation driven variability, stars modestly less massive than Sun (i.e., $M_{\star} \simeq~0.7 - 0.9M_\odot$, or K5-G8) appear to be better targets for EPRV surveys of Earth-analogs than stars somewhat more massive than the Sun (i.e., $M_{\star} \ge~1.2M_\odot$, or F7 and earlier). 
As seen in \autoref{sec:stellar_prop} and \autoref{fig:mass_plots}, the masses of planets around the lower mass stars will be measured with better fractional precision than Earth analogs around solar-mass stars, due to a combination of the increased signal as a result of the closer habitable zones (primary effect, $K \propto M^{-1.5}$) and the decreased granulation and oscillation amplitudes and timescales of their host stars (secondary effect $\sigma_{K} \propto M^{0.5}$).

\subsubsection{Target Selection}
In light of the strong dependence of planet mass SNR on stellar mass in the presence of correlated noise (as shown in \autoref{fig:mass_plots}), future survey simulations may wish to adjust the choice of target stars.

For example, it may be worthwhile to remove the small sample of high-mass stars where the planet mass SNR is unlikely to exceed $\sim$5 in a 15-year survey and reallocating that time to the remaining stars.
Alternatively, one could adjust the observing scheduler, so as to aim for a more uniform sensitivity across host star mass.
Implementing such design choices will depend on the exact survey goals, and will require future simulations to adequately test their impact.

\subsection{Caveats}\label{sec:disc:caveats}
\subsubsection{A Priori Covariance Matrix}\label{sec:disc:a_priori}
Our analysis has assumed that the covariance matrix for the GP noise models is known \emph{a priori}, and that the only free parameters are planet parameters.
In practice, the underlying covariance matrix will not be perfectly known, and will contribute additional loss of precision in the planet mass SNR. 
While the granulation and oscillation models are rooted in relatively well-established asteroseismic scalings, the AR model used here is intentionally a first-order approach. However, it is reasonable to assume that any star selected for such dedicated surveys will have additional observations that can help constrain the appropriate hyperparameters of a more accurate activity model (e.g., high-cadence photometry, knowledge of long-term activity cycles from legacy \ion{Ca}{2} H\&K measurements).

\subsubsection{Improve model for Active Regions \& Rotationally-linked Activity as a function of stellar properties}\label{sec:disc:activity_dependence}
As described in \autoref{sec:cov_kernels}, the active regions model we implement comes from a fit to simulated solar observations based on SOAP2.0 \citep{Dumusque2014} that also used simultaneous activity indicators.
Therefore, the model assumes a base level of activity mitigation and it is likely that these results represent a reduced active regions component, as opposed to the full expected solar variability due to active regions. 
Indeed, the short timescale and low amplitude indicate that this may indeed be the case, as it is dominated not by rotational modulation, but other shorter-timescale effects. 
However, the model is useful as it attempts to represent a realistic level down to which we may be able to correct for using current methods (i.e., using simultaneous photometry, or removing correlations with activity indicators) rather than an optimistic or pessimistic scenario. 
Recently, \citet{Langellier2020} used a quasi-periodic kernel fit to 800 days of HARPS-N solar data to describe the variability of activity-induced signals, which provides a ground-truth test for the effects of activity-induced variability. 

Both of these models for stellar variability \citep{Gilbertson2020,Langellier2020} are based on  the Sun.  Therefore, the amplitudes and timescales for the covariance kernels do not depend on stellar properties.
A more sophisticated model that scales the kernel parameters based on stellar properties is beyond the scope of this paper.
For now, we estimate the effect such a model might introduce on our results.  
First, consider the sign of the mass dependence of the white noise amplitude of activity-induced variability due to faculae/plages/spots. 
Among F, G, and K stars, activity-induced RV variability is positively correlated with mass \citep[e.g.,][]{Wright2004,Isaacson2010,Luhn2020a}. 
A simple model where the amplitudes, $\{a_0,a_1\}$ in \autoref{eqn:spots_kernel}, scale with stellar mass (or $\Teff$) would therefore introduce an additional positive mass dependence for $\sigma_{K}$.
As a result, we would expect steeper negative slopes in \autoref{fig:mass_plots}, with their value at $M_{\star} = M_{\odot}$ fixed. 
Such a model would reduce the activity-induced amplitudes of the low mass stars in the sample, and thus increase the threshold stellar mass at which Earth-analogs are no longer characterized with planet mass SNR $>$ 10.
Another simplified approach would be to scale the amplitudes with measured activity metrics, such as $\sigma_{K} \propto {R'_{HK}}^{1.66}$ as in \citet{Gupta2021}.
\citet{Haywood2020} found that the unsigned magnetic flux of the Sun is a good proxy for rotationally-modulated activity-induced velocity variations. Unfortunately, it has not yet been demonstrated that this it is practical to measure this quantity for other stars.
If it becomes practical for measures of the unsigned magnetic flux to be extended to other stars, then the results could inform the activity model in future EPRV survey simulations and/or lead to improved activity mitigation strategies.

In addition to amplitude scalings, we must also account for dependence of the correlation timescales on stellar properties.
It is reasonable to expect that the correlation timescale would scale with the stellar rotation period \citep[e.g.][]{Giles2017}. 
As a population, more massive stars rotate more rapidly than less massive star \citep{VanSaders2013}.
Since shortening the correlation timescale has the effect of increasing the planet mass SNR, this effect is expected to partially counteract the effect of scaling the amplitude with stellar mass.

\subsubsection{Magnetic Activity Cycles}\label{sec:disc:cycles}
We have not considered long term activity trends in this analysis. 
It is currently impractical to implement a realistic global population model that accounts for long period activity cycles.
Legacy surveys like the Mt. Wilson survey \citep[e.g.,][]{Baliunas1995} and the California Planet Search \citep[e.g.,][]{Isaacson2010} have monitored the \ion{Ca}{2} H \& K activity for stars spanning several decades.
However, while recent efforts have provided long time baseline analyses of activity cycles \citep{Baum2022}, clear trends between stellar properties and magnetic cycle periods and amplitudes and stellar properties have not been established. 
Further, it is not possible to predict \emph{a priori} whether a star will exhibit cyclical behavior in its activity time series.
Even among known cycling stars, it is not possible to predict the amplitude---or even the sign---of the corresponding RV effect.
Since long term activity cycles would be well sampled both in RV's and various simultaneous activity metrics, one could model them with an additional GP term.  
However, we expect that the effects of activity cycles are more complex.  
For example, the amplitude of variability induced by active regions would increase during periods of high activity and vice versa \citep{Meunier2010}.
This could be modeled hierarchically, e.g., one GP with a long correlation timescale to model the magnetic activity cycle that multiplies the GP for active regions (\autoref{eqn:spots_kernel}).
We leave the development of such a model to future studies.

\subsubsection{Effects of unknown orbital period, phase, non-zero eccentricities, and inclination effects}
\label{sec:disc:periods}
Our calculations have assumed circular orbits with known orbital period and phase.  
In practice, EPRV surveys uncertainties in the orbit will result in covariances between planet masses and orbit parameters and necessitate more computationally expensive analyses (e.g., MCMC) to characterize the uncertainties in planet masses and orbits.  
Given the number of observations in EPRV surveys and the orbital periods of habitable zone planets, we anticipate that aliases due to the solar day and lunar month be much less problematic than they can be for planets with shorter orbital periods.  
However, we expect added difficulties for planets with an orbital period close to one solar year, particularly for stars far from the ecliptic poles (i.e., subject to annual observability constraints).  
For such planets, the survey duration may need to be extended for a timescale greater than $\simeq \left|P^{-1} - (\mathrm{solar\, year})^{-1} \right|^{-1}$ in order to sample the full range of orbital phases and build confidence that the putative RV signal is not an artifact.  

We have assumed edge-on orbits, as done in \citep{EPRVWGreport}, so as to allow direct comparisons.
For understanding the effect of correlated noise, assuming an edge-on orbit is sufficient. 
However, the SNR that we \citep[and][]{EPRVWGreport} report overestimates the signal expected with these noise models for a population of planets with randomly oriented orbits. 
When choosing how much observing time to allocate to reach a desired SNR threshold, this effect should be considered.  
To account for a realistic inclination distribution, the SNR values reported here can be multiplied by $\sin i$, where $\cos i$ is drawn from a uniform distribution. Assuming randomly oriented orbital planes, the median of $\sin{i}$ $\sim$0.866. 
For our baseline survey simulations, assuming isotropic orbit orientations would decrease the percentage of planets with masses measured with SNR~$>$~10 in \IIa{} from 17\% to 7\% and percentage of planets with masses measured with SNR~$>$~5 from 81\% to 71\%.

\subsubsection{Effects of additional planets}
Our calculations have assumed each stars hosts a single planet.  
In practice, most inner planetary systems contain multiple planets \citep{He2020}.  
Inferring the number of planets in a system from RV observation is a long-standing challenge \citep[e.g.,][]{Nelson2014}.  
Even once the number and approximate orbital periods of planets are known, inferring the masses and orbits of multiple planets simultaneously is expected to result in increased uncertainties and sometimes non-trivial correlations in the their masses and orbital parameters.  
Near-degeneracies can be particularly problematic for planets with period ratios near 1:2 (or other low-order mean motion resonances).   
While the fraction of planetary systems with dynamically significant mean motion resonances is relatively small \citep{VerasFord2012}, planets near the 1:2 (and 2:3) mean-motion resonances are relatively common \citep{Fabrycky2014}.  We encourage future EPRV survey simulations to consider the implications of multiple planet systems.

\subsection{Opportunities for Further Research}
\subsubsection{Incorporating Recent Wavelength-Domain Data Analysis Strategies for Mitigating Stellar Variability}
Recent wavelength-domain efforts have also focused on disentangling spectral line deformations due to variability from Doppler shifts due to center-of-mass motions from orbiting planets \citep[e.g.,][]{Davis2017,Dumusque2018,Wise2018,CollierCameron2020,Zhao2020}. 
Such techniques may result in the ability to ``clean" the RV time series, removing some level of variability. 
While these techniques focus specifically on activity-induced variability, they may be more broadly applicable to any spectral deformation, for example, those from granulation or oscillations. 
However, these methods have yet to be tested for the observing cadence or long survey durations expected for the types of surveys studied in this work.
It is possible that improvements in stellar variability mitigation will result in a simple amplitude reduction for each of the active regions, granulation, and oscillation components studied here. However, it is also possible that the residuals after applying such techniques will have a different correlation structure.
In either case, the correlated noise models in this work should be updated as the RV community continues to improve their ability to disentangle Doppler shifts and stellar variability.

\subsubsection{Considering Alternative Observing Strategies for Mitigating Stellar Variability}
In this study, we calculated the impact of correlated noise on the planet mass SNR for one reasonable survey strategy.  Future studies could investigate alternative observing strategies.  For example, a survey could intentionally take multiple observations of the same star each night to help reduce the effects of granulation.  As another example, a survey might intentionally clump observations of a given target, so as to better characterize rotationally-linked variability (e.g., intensive observations over 3 rotation periods bracketed by observations at a lower rate.)

\subsubsection{Empirical measurements of Activity \& \texorpdfstring{$\nu_{max}$}{numax}}
In this work, we assumed all stars have the same level of activity. In practice, each star will differ in activity, which can be constrained with some combination of \emph{a priori} knowledge (e.g., through legacy \ion{Ca}{2} H \& K surveys like the Mt. Wilson program) and simultaneous \ion{Ca}{2} emission monitoring, as recommended by \citet{Crass2021}.  
In \autoref{sec:disc:activity_dependence}, we discussed how our results might be affected by including an activity component that depends on stellar properties. 
It is also possible that an activity noise model that depends on empirical activity indicators would shed light on the survey target selection.
Incorporating empirical measurements of activity might then allow surveys to target only the stars for which we expect to measure planet masses of Earth analogs to better than 20\%, (i.e., removing the high-mass end of \autoref{fig:mass_plots}) and allow the observation time to be given back to the remaining sample.

We have also used scaling relations for $\nu_{\max}$ that depend on $\logg$ and $\Teff$. In principle, one could measure the $\nu_{\max}$ for every star on the target list, so as to include a more accurate description of that star's oscillations and granulation spectrum. 
A direct measurement of $\nu_{\max}$ would reduce the uncertainty associated with the granulation and oscillation amplitudes and timescales, which will otherwise be driven by uncertainty in estimating $\logg$.
We expect that more precise measurements of $\nu_{\max}$ will improve the accuracy and precision of the calculated planet mass SNR for each star by only a small amount, given that the contribution of granulation and oscillations overall is small compared to that of rotationally linked activity.
While it is not likely that such precision will affect the results for this type of study, such efforts may prove worthwhile for properly analyzing and interpreting the data from these RV surveys on a star by star basis.

\subsubsection{Statistical methodology for detecting planets in presence of stellar variability}
In this work, we've focused on what the formal measurement precision for K would be.
That's similar to, but different than actually detecting a planet.
Things get even more complicated when there are likely multiple planets, each of which could be aliased with each other, stellar activity, window function, etc.
Doing proper planet detection calculations would be much more computationally expensive and the formal measurement precision calculated here serves as a basic check on our expectations for various survey designs.

We have also assumed that we’re working on a univariate time series of velocity measurements that have been cleaned of any contamination due to stellar variability. 
In practice, RV surveys will produce a multivariate time series including a raw RV measurement and multiple stellar variability indicators that also provide information about the state of any active regions, and potentially granulation and oscillations.
Future studies should explicitly consider the how uncertainty in the contribution from stellar variability propagates to affect the accuracy and precision of planet mass measurements.
When there is a potentially viable/practical strategy for using those, then it will be necessary to rerun simulations to provide more realistic results.  
In this paper, we have explored solving the problem of stellar variability by simply getting lots of observations with current accuracy.
There are complementary efforts to try to measure velocities in ways that are less contaminated by stellar variability that may provide additional improvements \citep[e.g.,][]{Davis2017,Dumusque2018,Wise2018,CollierCameron2020,deBeurs2020,Holzer2021,Cretignier2021}.

\section{Summary \& Conclusions}\label{sec:summary}
We provide easy-to-use Gaussian process (GP) models for the apparent radial velocity (RV) signals caused by stellar oscillations and granulation as a function of stellar properties.  
We provide analytic expressions that also account for effects of finite integration time. 
active regions (AR) and rotationally linked stellar variability \citep{Gilbertson2020}, we have a powerful toolbox for quantifying the effects of stellar variability on EPRV surveys.

We find that accounting for correlated noise is critical to accurately estimating the mass precision of future EPRV surveys. 
The effect of correlated noise reduces the planet mass SNR of Earth analogs by a factor of 7.42 from the no variability case and a factor of 1.86 from the white noise case for the survey design that performs best, \IIa.

Rotationally linked stellar variability has the largest effect (compared to granulation or oscillations) on the detection and characterization of Earth-analogs in EPRV surveys. 
Our AR-only model with correlated noise results in a reduction of the planet mass SNR of Earth analogs by a factor of 6.64 from the no variability model and a factor of 1.67 from our AR-only white noise model for \IIa.

Granulation also has a significant effect on the detection and characterization of Earth-analogs, particularly for stars more massive than the Sun.
Our granulation-only model with correlated noise results in a reduction of the planet mass SNR of Earth analogs by a factor of 2.2 from our granulation-only white noise model and a factor of 2.5 from the no variability model for \IIa.

The current observational strategy for mitigating the effects of oscillations on RV measurements is sufficient to characterize the masses of Earth-analogs.
Our oscillation-only model with correlated noise results in a reduction of the planet mass SNR of Earth analogs by a factor of 1.5 from our oscillation-only white noise model and a factor of 1.5 from the no variability model for \IIa.

The choice of observatory architecture significantly affects the precision of mass estimates and the fraction of Earth-analog planets which are expected to have masses measured with a given accuracy. 
For the default 10-year surveys with 5~\cms{} instrumental precision, the percentage of Earth analogs with planet masses measured with SNR $>$ 10 varies from 7\% to 17\% (4\% to 7\% when inclination is random) among the architectures studied.

Increasing the survey duration from 10 to 15 years (and thus increasing the number of observations by 50\%) can significantly affect the science yield of future EPRV surveys such as those proposed by \citet{EPRVWGreport}.
For 15-year surveys with 5~\cms{} instrumental precision, the percentage of Earth analogs with planet masses measured with SNR $>$ 10 roughly doubles in each architecture compared to the 10-year survey, varying from 21\% to 34\% among the architectures studied. As these improvements follow the expectations of $\sqrt{N}$, they arise primarily from the increased number of observations during the longer survey duration. A 10-year survey with 50\% more observations may see similar results, however the increased rate of observations will increase the correlations between observations, which may affect each architecture differently. Future survey simulations may wish to consider such an approach.

Improving instrumental precision (or reducing non-calibrated instrument noise) from 30~\cms{} to 10~\cms{} is expected to nearly double the percentage of planet masses measured with precision better than 10\% (or an increase of 8-10 percentage points for mass precision better than 20\% ).

Improving to the instrumental precision from 10~\cms{} to 5~\cms{} is not expected to have a significant impact on the yield of EPRV surveys similar to those proposed by \citet{EPRVWGreport}.

Our simulations indicate that none of the five survey architectures proposed by \citet{EPRVWGreport} studied in this work would be expected to measure masses of Earth-analogs with 10\% precision (or better) for most  of the host stars on their ``green target list,'' given the assumptions and caveats as detailed in \autoref{sec:disc:caveats}.

Our simulations suggest all five of the survey architectures proposed by \citet{EPRVWGreport} studied in this work (\I, \IIa{}, \IIb{}, \VIIIa{}, and \VIIIb) are expected to measure masses of Earth-analogs with 20\% precision (or better) for $>$70\% of the host stars on their ``green target list''.

Our simulations suggest that three of the five survey architectures proposed by \citet{EPRVWGreport} (\IIa{}, \IIb{}, and \VIIIb) could be expected to measure masses of Earth-analogs with 10\% precision (or better) for $>$30\% of the host stars on their ``green target list'', if only the survey duration were extended from 10 to 15 years.

We recommend future research to better characterize the magnitude and temporal characteristics of apparent RV variability due to faculae, star spots, and other rotationally-linked stellar variability, particularly for stars likely to be targeted by future direct imaging mission---as also suggested in Table A-3 of \citet{Crass2021}.

As researchers develop various methods for modeling and mitigating the effects of stellar variability, it will be important to characterize both the magnitude and the temporal correlations of the residuals (i.e., the apparent stellar RV that is not subtracted by the model) for each potential mitigation method.  When assessing the utility of potential strategies for mitigating stellar variability, it will be important to use simulations such as in this study to quantify the benefit of each method (as opposed to simply reporting the magnitude of residuals).  

\begin{acknowledgments}
We thank the anonymous referee, as well as Megan Bedell, Jason Wright, and Suvrath Mahadevan for their helpful comments and suggestions.
This research was partially supported by Heising-Simons Foundation Grant \#2019-1177. 
This work was partially supported by NASA Exoplanet Research Program Grant \#80NSSC18K0443.
This work was supported by a grant from the Simons Foundation/SFARI (675601, E.B.F.).
This work was partially supported by funding from the Center for Exoplanets and Habitable Worlds, which is supported by the Pennsylvania State University, the Eberly College of Science, and the Pennsylvania Space Grant Consortium.

Part of this research was carried out at the Jet Propulsion Laboratory, California Institute of Technology, under a contract with the National Aeronautics and Space Administration (NASA).

This research has made use of NASA's Astrophysics Data System Bibliographic Services. This material is based upon work supported by the National Science Foundation Graduate Research Fellowship Program under Grant No. DGE1255832.
\end{acknowledgments}

\bibliography{library}

\appendix
\section{Integrating the Oscillation Kernel}\label{sec:osc_derivation}
Here we describe the treatment of the double integral of the oscillation kernel performed for two finite exposures. However, we first notice that because $\Delta \equiv \left|t-t'\right|$, when two observations overlap t-t' becomes negative, and we must integrate the inverse function (i.e., $\Delta \rightarrow -\Delta$). While it is impossible for a single telescope to have 2 overlapping observations, the telescope architectures in this work may result in occasional overlapping observations. More importantly, overlap \emph{will} occur for a single telescope when computing the covariance matrix diagonal elements $(i.e.,\Delta=0)$. We therefore compute the double integral twice: in the case of two non-overlapping observations and two completely overlapping observations. Any two observations can then be broken up into a sum of various completely overlapping and non-overlapping segments.

\subsection{Non-overlapping observations}
In the case of non-overlapping observations ($\Delta > \delta_1/2 + \delta_2/2$), the double integral is 
\begin{equation}
k_{osc,sep}\left(\delta_1,\delta_2,\Delta\right) = \frac{1}{\delta_1 \delta_2} \int_0^{\delta_1}dt' \int_{\frac{\delta_1-\delta_2}{2}+\Delta}^{\frac{\delta_1+\delta2}{2}+\Delta} k_{osc}(t,t')\, dt
\label{eqn:osc_double_integral}
\end{equation}
where the term $k_{osc}(t,t')$ is simply \autoref{eqn:covar_osc_inst} with $\Delta$ replaced by $(t-t')$, where we have removed the absolute value under the assumption $t > t'$ for the two observations. Note that we have assumed a constant photon arrival time by dividing by $\delta_1 \delta_2$. While in practice this will not always be the case, it is not expected that the photon arrival rate will vary significantly over the typical exposure duration. 
We show the result of this integration here, first by defining the functions
\begin{equation}
I_1\left(y_i,y_j\right) = \frac{e^{-ay}\left(1-a^2\right)}{\eta \omega_{osc} \left(1+a^2\right)} \left(\cos{y} + a\sin{y} \right) \Big|_{y=y_i}^{y=y_j}
\label{eqn:I1}
\end{equation}
and
\begin{equation}
I_2\left(y_i,y_j\right) = -\frac{2a e^{-ay}}{\eta \omega_{osc} \left(1+a^2\right)} \left(\sin{y} - a\cos{y} \right) \Big|_{y=y_i}^{y=y_j}
\label{eqn:I2}
\end{equation}

where $a = 1/(2 Q \eta)$. The double integral in \autoref{eqn:osc_double_integral}---the covariance between the two non-overlapping observations---is then
\begin{equation}
\begin{split}
k_{osc,sep}\left(\delta_1,\delta_2,\Delta\right) =  \frac{S_0 Q }{\delta_1 \delta_2 \eta \left(1+a^2\right)} \Big(I_1\left(y_1,y_2\right) - I_2\left(y_1,y_2\right) 
- I_1\left(y_3,y_4\right) + I_2\left(y_3,y_4\right)\Big) 
\label{eqn:osc_sep}
\end{split}
\end{equation}
where 
$$y_1 = \eta \omega_{osc} \left(\frac{\delta_1 + \delta 2}{2} + \Delta\right),$$
$$y_2 = \eta \omega_{osc} \left(\frac{\delta_1 + \delta 2}{2} + \Delta - \delta_1\right), $$
$$y_3 = \eta \omega_{osc} \left(\frac{\delta_1 - \delta 2}{2} + \Delta\right), $$
$$y_4 = \eta \omega_{osc} \left(\frac{\delta_1 - \delta 2}{2} + \Delta -\delta_1\right), $$

\subsection{Completely overlapping observations}
In the case of completely overlapping observations, $\delta_1 = \delta_2 \equiv \delta$, $\Delta=0$, and \autoref{eqn:osc_double_integral} becomes
\begin{equation}
k_{osc,overlap}\left(\delta\right) = \frac{1}{\delta^2}\int_{0}^{\delta}dt' \int_{0}^{\delta} k_{osc}(t,t') \, dt
\label{eqn:osc_double_integral_overlap}
\end{equation}

For this integral, we see that because $\Delta$ in \autoref{eqn:covar_osc_inst} is equal to $\left|t-t'\right|$, the integrand is different when $t-t' < 0$. We can split the integral into positive and negative segment, but note that the positive segment is simply two times the negative segment. We show the result here
\begin{equation}
\begin{split}
k_{osc,overlap}\left(\delta\right) =  \frac{2 S_0 Q}{\delta^2 \eta \left(1+a^2\right)} \Big(I_1\left(y_1,y_2\right) - I_2\left(y_1,y_2\right) 
+ 2 a \delta \Big) 
\label{eqn:osc_over}
\end{split}
\end{equation}
with the same $I_1(y_i,y_j)$ and $I_2(y_i,y_j)$ defined above in \autoref{eqn:I1} and \autoref{eqn:I2} and with the same $y_1$ and $y_2$ as above, which evaluate to
$$y_1 = \eta \omega_{osc} \delta,$$
$$y_2 = 0$$
For any pair of observations, $\delta_1$, $\delta_2$, and $\Delta$, we can break up the result into segments of overlapping (\autoref{eqn:osc_over}) and nonoverlapping (\autoref{eqn:osc_sep}) portions to calculate the covariance. \autoref{fig:osc_kernel} shows the covariance as a function of $\Delta$ for various exposure lengths $\delta_1 = \delta_2$, assuming solar parameters. 

\section{Integrating the Granulation Kernel}\label{sec:gran_derivation}
We follow the same approach as in \autoref{sec:osc_derivation} but for the granulation kernel given in \autoref{eqn:gran_covar}

\subsection{Non-overlapping observations}
The double integral in this case is identical to \autoref{eqn:osc_double_integral} but replacing $k_{osc}(t,t')$ with $k_{gran}(t,t')$ given in \autoref{eqn:gran_covar}. The integral can be performed by parts and we show the result here, by first defining the new function
\begin{equation}
I\left(y_i,y_j\right) = e^{-y_i}\sin{y_i} - e^{-y_j}\sin{y_j}
\end{equation}
so that the final covariance is 
\begin{equation}
\begin{split}
k_{gran,sep}\left(\delta_1,\delta_2,\Delta \right) =  \frac{S_1}{\omega_1 \delta_1 \delta_2} e^{-\frac{\pi}{4}}\Big(I(y_{13},y_{14}) - I(y_{11},y_{12})\Big) + \frac{S_2}{\omega_2 \delta_1 \delta_2} e^{-\frac{\pi}{4}}\Big(I(y_{23},y_{24}) - I(y_{21},y_{22})\Big)
\label{eqn:gran_sep}
\end{split}
\end{equation}
where 
$$y_{11} = \frac{\omega_1}{\sqrt{2}} \left(\frac{\delta_1 + \delta_2}{2} + \Delta\right) - \frac{\pi}{4},$$
$$y_{12} =  \frac{\omega_1}{\sqrt{2}} \left(\frac{\delta_1 + \delta_2}{2} + \Delta - \delta_1\right)- \frac{\pi}{4}, $$
$$y_{13} = \frac{\omega_1}{\sqrt{2}} \left(\frac{\delta_1 - \delta_2}{2} + \Delta\right)- \frac{\pi}{4}, $$
$$y_{14} = \frac{\omega_1}{\sqrt{2}} \left(\frac{\delta_1 - \delta_2}{2} + \Delta -\delta_1\right)- \frac{\pi}{4}, $$
$$y_{21} = \frac{\omega_2}{\sqrt{2}} \left(\frac{\delta_1 + \delta_2}{2} + \Delta\right) - \frac{\pi}{4},$$
$$y_{22} =  \frac{\omega_2}{\sqrt{2}} \left(\frac{\delta_1 + \delta_2}{2} + \Delta - \delta_1\right)- \frac{\pi}{4}, $$
$$y_{23} = \frac{\omega_2}{\sqrt{2}} \left(\frac{\delta_1 - \delta_2}{2} + \Delta\right)- \frac{\pi}{4}, $$
$$y_{24} = \frac{\omega_2}{\sqrt{2}} \left(\frac{\delta_1 - \delta_2}{2} + \Delta -\delta_1\right)- \frac{\pi}{4}, $$

\subsection{Overlapping observations}
As with the oscillation kernel, in the case of completely overlapping observations, the double integral is given in \autoref{eqn:osc_double_integral_overlap}, but now using \autoref{eqn:gran_covar} in the integrand. We can again split the integral into a positive and negative segment, choosing to evaluate the positive segment and multiplying by a factor of 2. We show the result here
\begin{equation}
k_{gran,overlap}(\delta) = \frac{2S_1}{\omega_1 \delta^2} e^{-\frac{\pi}{4}}\Big(I(y_{12},y_{11}) \Big) + \frac{2S_1}{\delta} +  \frac{2S_2}{\omega_2 \delta^2} e^{-\frac{\pi}{4}}\Big(I(y_{22},y_{21}) \Big) + \frac{2S_2}{\delta}
\label{eqn:gran_over}
\end{equation}
where $y_{11},y_{12},y_{21}$ and $y_{22}$ are the same as above but simplified under the assumption $\delta_1 = \delta_2 \equiv \delta$, $\Delta=0$

$$y_{11} = \frac{\omega_1}{\sqrt{2}} \delta - \frac{\pi}{4},$$
$$y_{12} =  - \frac{\pi}{4}, $$
$$y_{21} = \frac{\omega_2}{\sqrt{2}} \delta - \frac{\pi}{4},$$
$$y_{22} =  - \frac{\pi}{4}, $$

\end{document}